\documentclass[11pt]{article}
\usepackage{epsfig,cite}
\usepackage{amsmath}
\usepackage{amssymb}
\usepackage{amsfonts}
\usepackage{latexsym}
\usepackage{wasysym}
\usepackage{bm}
\def\be{\begin{equation}}
\def\ee{\end{equation}}
\def\bea{\begin{eqnarray}}
\def\eea{\end{eqnarray}}

\def\smallfrac#1#2{\hbox{${\frac{#1}{#2}}$}}
\newcommand{\bi}{\begin{itemize}}
\newcommand{\ei}{\end{itemize}}
\newcommand{\ben}{\begin{enumerate}}
\newcommand{\een}{\end{enumerate}}

\newcommand{\lp}{\left(}

\newcommand{\rp}{\right)}

\def\frac#1#2{{{#1}\over {#2}}}
\def\half{\smallfrac{1}{2}}
\def\gsim{\mathrel{\rlap{\lower4pt\hbox{\hskip1pt$\sim$}}
    \raise1pt\hbox{$>$}}}         %greater than or approx. symbol
\def\lsim{\mathrel{\rlap{\lower4pt\hbox{\hskip1pt$\sim$}}
    \raise1pt\hbox{$<$}}}         %less than or approx. symbol

\newcommand{\cov}{\mathrm{cov}}

\newcommand{\draft}[1]{}

\def\ev#1{\langle #1 \rangle}
\def\cov{{\rm cov}}
\def\covinv{{\rm cov}^{-1}}
\def\covz{{\rm cov}_0}

\def\covm{{\rm cov}_m}
\def\covminv{{\rm cov}_m^{-1}}
\def\covt{{\rm cov}_t}
\def\covtinv{{\rm cov}_t^{-1}}
\def\covtz{{\rm cov}_{t_0}}
\def\covtzinv{{\rm cov}_{t_0}^{-1}}

\usepackage{color}
\definecolor{comment}{rgb}{0,0.3,0}
\definecolor{identifier}{rgb}{0.0,0,0.3}

\usepackage{listings}
\begin{document}

\pagestyle{empty}

\begin{flushright}

Edinburgh-2009/22\\ 
IFUM-950-FT\\
FREIBURG-PHENO-09/09\\
CP3-09-51\\ 
\end{flushright}

\begin{center}
\vspace*{.2cm}
{\Large \bf Fitting Parton Distribution Data with 
Multiplicative Normalization Uncertainties}
 \\
\vspace{0.8cm}

{\bf  The NNPDF Collaboration:}\\
Richard~D.~Ball$^{1}$,
 Luigi~Del~Debbio$^1$, Stefano~Forte$^2$, Alberto~Guffanti$^3$,\\ 
Jos\'e~I.~Latorre$^4$, 
Juan~Rojo$^2$ and Maria~Ubiali$^{1,5}$.

\vspace{1.cm}
{\it ~$^1$ School of Physics and Astronomy, University of Edinburgh,\\
JCMB, KB, Mayfield Rd, Edinburgh EH9 3JZ, Scotland\\
~$^2$ Dipartimento di Fisica, Universit\`a di Milano and
INFN, Sezione di Milano,\\ Via Celoria 16, I-20133 Milano, Italy\\
~$^3$  Physikalisches Institut, Albert-Ludwigs-Universit\"at Freiburg,
\\ Hermann-Herder-Stra\ss e 3, D-79104 Freiburg i. B., Germany  \\
~$^4$ Departament d'Estructura i Constituents de la Mat\`eria, 
Universitat de Barcelona,\\ Diagonal 647, E-08028 Barcelona, Spain\\
~$^5$Center for Particle Physics and Phenomenology (CP3),\\ Universit\'e Catholique de Louvain,\\ Chemin du Cyclotron, B-1348 Louvain-la-Neuve, Belgium\\ }
%\End{center}
\bigskip
\bigskip

{\bf Abstract}
\end{center}
\noindent
The extraction of robust parton distribution functions with faithful
errors requires a careful treatment of the uncertainties in the
experimental results. In particular, the data sets used in current
analyses each have a different overall multiplicative normalization 
uncertainty that needs to be properly accounted for in the fitting procedure.  
Here we consider the generic problem of performing a global fit to many
independent data sets each with a different overall multiplicative
normalization uncertainty. We show that the methods in common use to
treat multiplicative uncertainties lead to systematic biases. We
develop a method which is unbiased, based on a self--consistent
iterative procedure.  We then apply our generic method to the determination
of parton distribution functions with the NNPDF methodology, which
uses a Monte Carlo method for uncertainty estimation.

\vspace*{1cm}

\vfill
\noindent

\begin{flushleft} November 2009 \end{flushleft}
\eject

\setcounter{page}{1} \pagestyle{plain}

\tableofcontents

\section{Introduction}\label{sec:intro}

The interpretation of forthcoming experiments at the Large Hadron
Collider requires the development of precision statistical analysis
tools. One context where this is especially apparent is the
determination of the parton distribution functions of the proton,
which are obtained by global analysis of existing data sets
(see~\cite{HERALHC} for a review). These PDFs, together with their
associated uncertainties, should be without theoretical prejudice, and
should be associated with a genuine statistical confidence level. In a
series of recent
papers~\cite{NNPDFSF1,NNPDFSF2,NNPDFNS,NNPDF10,NNPDF11,NNPDF12} the
NNPDF collaboration has adopted a method based on a Monte Carlo 
estimate of uncertainties~\cite{Giele:2001mr} that allows one to
propagate the uncertainty on the experimental data to the fitted PDFs,
and to any other quantity which depends on them. The PDFs are
parametrized by neural networks, containing large numbers of free
parameters, sufficient to ensure that the resulting ensembles of
fitted PDFs are free from any bias due to assumptions about the
underlying functional form. Within this context, in which one aims
typically at accuracies at the percent level in physical observables,
more subtle problems in data analysis become relevant. One such
problem is the treatment of normalization uncertainties.

When combining data sets from independent experiments it is necessary 
to take account of the overall normalization uncertainty associated 
with each experiment: an experiment with large normalization 
uncertainty should contribute less to the fit than one with a 
small uncertainty. Normalization uncertainties are usually 
multiplicative, in the sense that each data point within the set has
a normalization uncertainty proportional to the measurement at that point.
All these normalization uncertainties are however correlated across the 
whole set of data points. Fitting the data with the usual Hessian 
method, using the complete covariance matrix, leads to a substantial 
bias in the fitted value due to the fact that smaller data 
points are assigned a 
smaller uncertainty than larger ones~\cite{dagos}. This problem is usually 
avoided by using the ``penalty trick'': the normalization of each data set 
is treated as a free parameter to be determined during the fitting, 
within a range restricted by the quoted experimental uncertainty. While 
this method indeed gives correct results when fitting data from a single 
experiment, we will show below that it remains biased when used in  
fits which combine several different data sets. 

In this paper we develop a treatment of normalization uncertainties which
is always free from bias, even when fitting to many different data
sets. While the original motivation of this study comes from PDF
determination, all our results are of general relevance whenever
a quantity has to be determined from data which are affected by
multiplicative uncertainties. Hence most of our discussion will be
completely general, and we will only consider the issue of PDF
determination in the end, as an example of the application of our
proposed method.

The paper is 
constructed as follows: in Sect.~2 we review 
the Hessian and Monte Carlo methods for the extraction of a quantity
from a set of experimental measurements, and how they might be implemented
when there are normalization uncertainties. In Sect.~3 we review the
well--known fact that
using the full covariance matrix for the treatment of normalization 
uncertainties leads to a bias for a single experiment, and we show that
 further biases arise when combining several experiments. 
In Sect.~4 we show that the so--called
``penalty trick'' method commonly used to overcome this bias, while working 
well for a single experiment, still leads to biased results when 
used to combine results from several experiments. In Sect.~5 we attempt 
to construct a self--consistent covariance matrix, but find that this 
too leads to biased results when used to combine several different 
experiments. In Sect.~6 we cure the defect in this method by determining 
the self--consistent covariance matrix through an iterative procedure 
using as a starting point the result of a previous fit: 
we show that this method (the ``$t_0$-method'') is 
completely unbiased and rapidly convergent.
Finally in Sect.~7 we discuss the treatment of normalization errors in
PDF fitting: after a brief summary of the methods used currently, we discuss
the effect of our new  $t_0$-method on the NNPDF1.2 parton
fit. Specifically, we demonstrate 
explicitly the practicality and convergence of the method, and
use it to quantify the effect of including the normalization uncertainties 
in the determination of PDFs.

The language of the paper is that of a particle physicist (and thus 
similar in approach to e.g. Ref.\cite{dagos}), not a professional statistician.
We thus make no pretence of mathematical rigour, but hope nonetheless that our 
results will be of practical use to physicists interested in fitting 
large datasets.

\section{Hessian and Monte Carlo}
\label{sec:HMC}

\subsection{Hessian Methods}
\label{subsec:Hess}

Consider a simple but typical experimental situation in which we have  
$n$ measurements $m_i$ of a single theoretical quantity $t$, with 
experimental uncertainties given by a covariance matrix $(\cov)_{ij}$:
typically this takes the form
\be
\label{eq:cov}
(\cov)_{ij}=\delta_{ij}\sigma_i^2 
+ \sum_{k=1}^n \bar\sigma_{ik}\bar\sigma_{kj},
\ee
where the $\sigma_i$ are uncorrelated uncertainties (typically
obtained as the sum in quadrature of statistical
and uncorrelated systematic uncertainties) and $\bar\sigma_{ik}$ are 
correlated (typically systematic)  uncertainties. All experimental 
uncertainties are generally assumed to be Gaussian.
Then the least squares estimate for $t$ is given by minimizing the $\chi^2$ 
function 
\be
\label{eq:chisq}
\chi^2(t) = \sum_{i,j=1}^n (t-m_i)(\covinv)_{ij}(t-m_j),
\ee
and thus by
\be
\label{eq:tcov}
t = \frac{\sum_{i,j=1}^n (\covinv)_{ij}m_j}{\sum_{i,j=1}^n 
(\covinv)_{ij}}.
\ee
The variance of $t$, $V_{tt}$ is found through
\be
\label{eq:vtt}
V_{tt} = \Big(\half\frac{\partial^2\chi^2}{\partial t^2}\Big)^{-1}
= \frac{1}{\sum_{i,j=1}^{n}(\covinv)_{ij}}.
\ee

%Note that this is a maximum 
%likelihood estimator: the likelihood 
%function $L(t) = N \exp(-\half \chi^2(t))$, where $N$ is a 
%normalization factor. Hence provided $N$ does not depend on
%the parameters which are minimized, as in this case, 
%maximum likelihood means minimization of the 
%$\chi^2(t)$. Maximum likelihood estimators are always consistent and 
%asymptotically unbiased (i.e. unbiased in the limit of large $n$).

Of course when there are several quantities $t$ to be determined, the
variances are given by the diagonal elements $V_{tt}$ of the matrix
determined by inversion of the Hessian matrix of second derivatives of
$\chi^2$: hence the name of the method.  Here the simplest case of
only one quantity $t$ will be sufficient to illustrate the points we
wish to make.  The situation is yet more complicated when fitting
parton distribution functions: then, $t$ is actually some nontrivial
but calculable function (such as a cross section or structure
function) of the (many) fitted parameters which describe the shape of
the underlying parton distribution functions which is being
determined. Again this complication is irrelevant to the issues to be
discussed here, so we will ignore it.

In the very simplest case in which the measurements $m_i$ have completely 
uncorrelated (e.g. purely statistical) uncertainties $\sigma_i$,
\be
\label{eq:covstat}
(\cov)_{ij} = (\covz)_{ij}=\sigma_i^2\delta_{ij},
\ee
so
\be
\label{eq:chisqstat}
\chi^2(t) = \sum_{i=1}^n \frac{(t-m_i)^2}{\sigma_i^2},
\ee
whence at the minimum $t=w$ and $V_{tt} = \Sigma^2$, where
\be
\label{def:w}
w \equiv \Sigma^2\sum_{i=1}^n \frac{m_i}{\sigma_i^2},
\ee
and $\Sigma^2$ is given by
\be
\label{def:Sig}
\frac{1}{\Sigma^2} = \sum_{i=1}^n \frac{1}{\sigma_i^2}.
\ee
Thus the theoretical quantity $t$ is given by the average value of the 
measurements $m_i$ weighted by the inverses of the variances $\sigma_i^2$, and 
the inverse of the variance of $t$ is 
likewise the average of the inverses of the variances $\sigma_i^2$. 

When all the variances are equal, $\sigma_i=\sigma$, and $w=\bar{m}$, where
\be
\label{def:mbar}
\bar{m} = \frac{1}{n}\sum_{i=1}^n m_i,
\ee
i.e. the unweighted average of the measurements. In this limiting
case, the determination Eq.~(\ref{eq:tcov}) of $t$ is manifestly seen
to be unbiased (and in particular it tends to the true value in the limit
of large number of measurements): no measurement is preferred, and
$\Sigma^2 = \sigma^2/n$, so the 
variance is reduced by a factor of $1/n$, due to there being $n$ independent 
measurements of the same quantity. In what follows we will use the words 
``biased'' and ``unbiased'' to describe estimates of the mean which pass or
fail this simple test. 

In the opposite extreme, if one of the variances, say $\sigma_i^2$, becomes 
very large compared to the others, the contribution of the measurement 
$m_i$ to $w$ and $\Sigma^2$ becomes very small, so this measurement decouples 
from the rest as it must. In what follows we shall use freedom from bias 
(as defined above) and decoupling as two criteria to assess the 
usefulness of a particular method of determining $t$ and its 
variance.\footnote{Note that the terms  
``bias'' and ``decoupling'' as used here are
not quite the same as the technical definitions of consistency and bias used by
statisticians, which distinguish more carefully between results 
obtained with finite size samples and those when the sample size 
becomes infinite.}

\subsection{The Monte Carlo Method}
\label{subsec:MC}

A different way of determining a theoretical quantity from a set of
measurements is to construct a Monte Carlo representation of the
data. First,  the $n$ data
points are associated to $n$ random variables $M_i$,
normally distributed around the averages $m_i$ according to the 
covariance matrix $(\cov)_{ij}$. Then, an ensemble  $\{M_i\}$ 
of replicas of the
data is constructed: these by construction satisfy
\be
\label{eq:mavg}
\ev{M_i} = m_i,\qquad \ev{M_i M_j} = m_im_j + (\cov)_{ij},
\ee
where $\langle\rangle$ denotes averaging over the set of replicas.
Finally
an ensemble $\{T\}$ of replicas of the theoretical quantity $t$ is
determined from  the data replicas $\{M_i\}$, by minimizing a suitable
error function $E_\mathrm{MC}(T)$.
The mean value and variance of $t$ (and indeed any other function of $t$) may 
then be found simply by averaging over the replicas: 
\be
\label{eq:mcres}
{\rm E}[t] = \ev{T},\quad
{\rm Var}[t] = \ev{T^2}-\ev{T}^2.
\ee

This method becomes advantageous when the determination of some
function of $t$ is called for: once the ensemble of replicas $\{T\}$
has been found, error propagation to any function of $t$, no matter
how complicated, may be performed by simply averaging over
replicas. This is especially useful in situations in which $t$ is
multidimensional, or a nontrivial function of some underlying
theoretical quantity (such as a PDF).

Clearly, the features of the result obtained with this method depend
on the choice of error function $E_\mathrm{MC}(T)$ which determines the
the ensemble of replicas $\{T\}$ from the data replicas  $\{M_i\}$.
In the simple situation discussed in the previous section 
we may choose $E_\mathrm{MC}(T)=\chi^2(T)$, where 
the $\chi^2$ function is given by Eq.~(\ref{eq:chisq}). Then
\be
\label{eq:Tcov}
T = \frac{\sum_{i,j=1}^n (\covinv)_{ij}M_j}{\sum_{i,j=1}^n 
(\covinv)_{ij}},
\ee
so that, using Eq.~(\ref{eq:mcres}),
\be
\label{eq:Et}
{\rm E}[t] = \frac{\sum_{i,j=1}^n (\covinv)_{ij}m_j}{\sum_{i,j=1}^n 
(\covinv)_{ij}},
\ee 
while
\bea
\label{eq:Vart}
{\rm Var}[t] 
&=& \frac{
\sum_{i,j,k,l=1}^n (\covinv)_{ij}(\covinv)_{kl}(\ev{M_jM_l}-\ev{M_j}\ev{M_l})}{\left(\sum_{i,j=1}^n (\covinv)_{ij}\right)^2}\nonumber\\
&=&\frac{1}{\sum_{i,j=1}^n (\covinv)_{ij}},
\eea 
in agreement with Eq.~(\ref{eq:tcov}) and Eq.~(\ref{eq:vtt}) found using 
the Hessian method.

\subsection{Normalization Uncertainties}
\label{subsec:Norm}

The two methods discussed in the previous two sections cover most 
situations of uncorrelated and correlated errors found in combining 
experimental data. However problems arise when data sets have 
overall multiplicative uncertainties, such as normalization
uncertainties:  this is
due to biases arising from 
the rescaling of errors~\cite{Cello,dagos94,dagos}. The effect of these 
biases in the naive application of the Hessian method can be very severe, 
as will be discussed in the following section. Here we consider 
normalization uncertainties in the Monte Carlo method, which is possibly more 
straightforward.

First consider the situation in which all the data come from a single 
experiment, with a single overall normalization uncertainty $s$, 
assumed to be Gaussian. 
In the Monte Carlo method the normalization uncertainty is taken into 
account by multiplying the data by a random
factor $N$, which is normally distributed around 1 with variance
$s$. Assuming $N$ to be uncorrelated with $M_i$
\be
\label{eq:navg}
\ev{N} = 1,\qquad \ev{N^2} = 1+s^2, \qquad 
\ev{f(N)g(M_i)}=\ev{f(N)}\ev{g(M_i)}.
\ee
When fitting to the replicas, the error function will depend 
on $NM_i$, but the weights of the different measurements will be unchanged, 
since an overall rescaling of the data should not affect 
the relative weight of the measurements in the fit. Thus we now take
\begin{equation}
  \label{eq:efNone}
  E_{\rm MC}(T) = \sum_{i,j=1}^n (T-NM_i)(\covinv)_{ij}(T-NM_j)
\end{equation}
where $(\cov)_{ij}$ is the same covariance matrix used when there was 
no normalization uncertainty.
The result of the minimization of Eq.~(\ref{eq:efNone}) 
is the same as Eq.~(\ref{eq:Tcov}), but 
with an overall factor of $N$ 
\be
\label{eq:TNcov}
T = N\frac{\sum_{i,j=1}^n (\covinv)_{ij}M_j}{\sum_{i,j=1}^n 
(\covinv)_{ij}},
\ee
so ${\rm E}[t]$ is the same as in Eq.~(\ref{eq:Et}), while now 
\bea
\label{eq:VartN}
{\rm Var}[t] &=& \frac
{\sum_{i,j,k,l=1}^n (\covinv)_{ij}(\covinv)_{kl}(\ev{N^2}\ev{M_jM_l}
-\ev{N}^2\ev{M_j}\ev{M_l})}{\left(\sum_{i,j=1}^n (\covinv)_{ij}\right)^2}\nonumber\\
&=&\frac{1+s^2}{\sum_{i,j=1}^n (\covinv)_{ij}}+ s^2 {\rm E}[t]^2.
\eea 
The first term on the right--hand side of 
Eq.~(\ref{eq:VartN}) is the same as was found previously in Eq.~(\ref{eq:Vart}),
but with an extra factor of $1+s^2$, while the second term is the 
contribution to the variance from the normalization 
uncertainty. This is as expected: indeed, the variance of a 
product of random variables is
\be
\label{eq:varnm}
{\rm Var}[NM]={\rm E}[N]^2{\rm Var}[M]+{\rm E}[M]^2{\rm Var}[N]
+{\rm Var}[N]{\rm Var}[M],
\ee
where the last term is usually neglected because, being a product of
two variances, it corresponds to a higher order moment of the
probability distribution for $NM$.

For the simple case of uncorrelated 
statistical measurement errors Eq.~(\ref{eq:covstat}), 
\begin{equation}
  \label{eq:efNonestat}
  E_{\rm MC}(T) = \sum_{i=1}^n \frac{(T-NM_i)^2}{\sigma_i^2},
\end{equation}
whence
\begin{equation}
  \label{eq:tminnsim}
  T = N \Sigma^2 \sum_{i=1}^{n}\frac {M_i}{\sigma_i^2},
\end{equation}
so ${\rm E}[t]=w$, Eq.~(\ref{def:w}), while 
\be
\label{eq:varnsim}
  {\rm Var}[t]= \Sigma^2 + s^2 w^2 + s^2\Sigma^2.  
\ee
Clearly then this method gives correct unbiased results for the case of a 
single experiment with normalization uncertainty.

Let us now consider a slightly more complex situation, where each of the 
measurements $m_i$ comes from a different experiment, 
and also has an independent normalization uncertainty $s_i$. Here and
henceforth when discussing this situation we will neglect possible
correlations between these measurements with independent
uncertainties, which are usually absent if the measurements are
obtained from independent experiments; however,  the inclusion of such
correlations is  straightforward and it does not affect our
subsequent results.
Following the same line of reasoning as above, we want to derive 
the fitted value for $t$ and its variance in the Monte Carlo approach. 
We thus introduce independent normally distributed random variables 
$N_i$ to represent the
normalization uncertainties, each with mean one, variance $s_i^2$, 
and uncorrelated to each other and to the $M_i$:
\be
\label{eq:navgN}
\ev{N_i} = 1,\qquad \ev{N_iN_j} = 1+s_i^2\delta_{ij}, \qquad 
\ev{f(N_i)g(M_j)}=\ev{f(N_i)}\ev{g(M_j)}.
\ee
The difficulty now is to choose an 
appropriate error function. The simplest choice would be
\begin{equation}
  \label{eq:efNoneNexp}
  E_{\rm MC}(T) = \sum_{i=1}^n \frac{(T-N_iM_i)^2}{\sigma_i^2},
\end{equation}
Minimizing with respect to $T$ and averaging over the replicas then 
gives ${\rm E}[t]=w$ as before, but 
\bea
\label{eq:VartNexp}{\rm Var}[t] &=& 
\Sigma^4\sum_{i,j=1}^n (\ev{N_iN_j}\ev{M_iM_j}
-\ev{N_i}\ev{N_j}\ev{M_i}\ev{M_j})/\sigma_i^2\sigma_j^2,\nonumber\\
&=&\Sigma^2 + \Sigma^4\sum_{i=1}^n s_i^2(m_i^2+\sigma_i^2)/\sigma_i^4.
\eea 
When all the experiments have the same normalization uncertainty, 
$s_i=s$, these results are not so unreasonable: 
${\rm E}[t]=w$, while 
\be
\label{eq:VartNexpeq}
{\rm Var}[t] = (1+s^2)\Sigma^2 + s^2 
\Sigma^4\sum_{i=1}^n \frac{m_i^2}{\sigma_i^4}\,.
\ee
However Eq.~(\ref{eq:efNoneNexp}) is clearly incorrect in general because 
differences in the normalization uncertainties 
$s_i$ are not taken account of in the weighting 
of the different measurements. In particular if one of the experiments 
has a relatively large normalization uncertainty, it still contributes 
to the mean, but spoils the measurement by giving a very large 
contribution to the variance. 

Therefore, results found using the Monte Carlo
method with the error function Eq.~(\ref{eq:efNoneNexp}) are unbiased 
when the normalization uncertainties are equal,
but do not satisfy the criterion of decoupling. 
It follows that when we have more than one experiment, and in particular 
when we wish to include experiments with a large overall normalization 
uncertainty, we need to choose 
a better error function than Eq.~(\ref{eq:efNoneNexp}), which incorporates 
differences in the normalization uncertainties $s_i$. We are thus led to 
consider error functions built using the full covariance matrix, 
including normalization uncertainties, and thus rather closer to the  
$\chi^2$-function.

\section{The d'Agostini Bias}
\label{sec:mcov}

We saw in the previous section that when we are combining different 
experiments with independent and different normalization uncertainties, 
it is necessary to incorporate these differences into the $\chi^2$-function 
or error function used in the fitting procedure. This is true both 
for the Hessian method and for the Monte Carlo method. In the previous
section we have shown that this is easily done in the case of a single
experiment, but when several experiments must be combined
the simplest choice of error function
Eq.~(\ref{eq:efNoneNexp})  leads to results which do not satisfy the
decoupling criterion. In this section 
we will see that in the Hessian case the simplest choice of error
function leads to results which are severely biased even in the case
of a single experiment.

Specifically, we consider the case in which normalization uncertainties
are included by using as an error function the $\chi^2$-function computed 
using the full covariance matrix including normalizations: 
\be
\label{eq:chisqm}
\chi^2_m(t) = \sum_{i,j=1}^n (t-m_i)(\covminv)_{ij}(t-m_j) .
\ee
As in the previous section we consider two cases separately: when we have 
$n$ measurements all made within a single experiment, and thus with a 
common normalization uncertainty, so
\be
\label{eq:covmoneexp}
(\covm)_{ij} = (\cov)_{ij}+s^2 m_im_j,
\ee
and then when each of the $n$ measurements is made in an independent 
experiment, all uncorrelated, and in particular with different 
normalization uncertainties, so   
\be
\label{eq:covmNexp}
(\covm)_{ij} = (\sigma_i^2+s_i^2 m_i^2)\delta_{ij}.
\ee
Note that the 
more realistic case in which there are $n_{\rm exp}$ experiments each 
with $n^i_{\rm dat}$ measurements can be built from these two simpler 
examples by first combining the many measurements in each individual 
experiment together into one measurement, using Eq.~(\ref{eq:covmoneexp}), 
and then combining the experiments using Eq.~(\ref{eq:covmNexp}), so these 
two cases should suffice to illustrate all the issues involved.

\subsection{One experiment}
\label{subsec:mcov1}

Consider first a very simple model of a single experiment with only 
two data points. The covariance matrix Eq.~(\ref{eq:covmoneexp}) is then simply
\begin{equation}
\label{covmoneexp2}
       (\covm)_{ij}=\left(\begin{array}{cc}\sigma_1^2 + s^2m_1^2&s^2m_1m_2\cr
                     s^2 m_1m_2& \sigma_2^2+s^2m_2^2\cr\end{array}\right)
\end {equation}
so the $\chi^2$ Eq.~(\ref{eq:chisqm}) is
\begin{equation}
  \label{eq:chi2oneexp}
  \chi_m^2(t) = \frac{(t-m_1)^2(\sigma_2^2+m_2^2 s^2) + 
(t-m_2)^2(\sigma_1^2+m_1^2 s^2) - 2(t-m_1)(t-m_2)m_1m_2s^2}
{\sigma_1^2\sigma_2^2 + (m_1^2\sigma_2^2+m_2^2\sigma_1^2)s^2}
  \, .
\end{equation}
Minimizing this $\chi^2$-function with respect to $t$ gives after a 
straightforward calculation the result
\begin{equation}
\label{eq:dagostini}
t = \frac{m_1/\sigma_1^2 + m_2/\sigma_2^2}
{1/\sigma_1^2 + 1/\sigma_2^2+(m_1-m_2)^2s^2/\sigma_1^2\sigma_2^2}
= \frac{w}{1+(m_1-m_2)^2s^2/\Sigma^2},
\end{equation}
where $w$ is the weighted mean Eq.~(\ref{def:w}) with $n=2$. 

It follows that when $m_1\neq m_2$ and $s\neq 0$ the result for $t$ has a 
downward shift. That this shift is clearly a bias can be seen for
instance by considering the simple case 
$\sigma_1=\sigma_2=\sigma$. Then 
Eq.~(\ref{eq:dagostini}) gives
\begin{equation}
\label{eq:dagostinieq}
t = \frac{\bar{m}}{1+2r^2s^2\bar{m}^2/\sigma^2}=
\bar{m}(1-2r^2s^2\bar{m}^2/\sigma^2 +O(r^4)).
\end{equation}
where we have defined
\be
\label{def:mrtwo} 
\bar{m}\equiv\half(m_1+m_2),\qquad r\equiv \frac{m_1-m_2}{m_1+m_2}.
\ee
Thus simply minimizing the $\chi^2$ derived from the correlated covariance 
matrix Eq.~(\ref{eq:covmoneexp}) leads to a central value which 
is shifted  downwards: for a sufficiently large $s^2/\sigma^2$ one
can get an average which is lower than either of the two values which
are being averaged, so one must conclude that the result is biased.

This bias gets worse as the number of data points 
increases~\cite{dagos94,dagos,Takeuchi:1995xe}. For 
$n$ data $m_i$, with statistical uncertainties $\sigma_i$, 
the covariance matrix given by Eq.~(\ref{eq:covmoneexp}) and 
Eq.~(\ref{eq:covstat}) has inverse
\be
\label{eq:ncovinv}
(\covminv)_{ij} =  
\frac{\delta_{ij}}{\sigma_i^2}-\frac{m_i m_j}{\sigma_i^2\sigma_j^2}
\frac{s^2}{1+s^2  m^2/\Sigma^2}.
\ee
where 
\be
m^2\equiv \Sigma^2 \sum_{i=1}^n \frac{m_i^2}{\sigma_i^2}.
\ee 
The $\chi^2$-function Eq.~(\ref{eq:chisqm}) is then minimized when 
\be
\label{eq:dagostinin}
t = \frac{\sum_{i,j=1}^n (\covminv)_{ij}m_j}{\sum_{i,j=1}^n 
(\covminv)_{ij}}
=\frac{w}{1+ r^2 s^2 w^2 /\Sigma^2}
\ee
where $w$ is defined in Eq.~(\ref{def:w}) while $r$ is defined through
\be
\label{def:r}
m^2-w^2 = \Sigma^2\sum_{i=1}^n \frac{(m_i-w)^2}{\sigma_i^2}
\equiv r^2 w^2.
\ee
So again we have a downwards bias unless $s^2 =0$ or all the measurements 
$m_i$ are equal.
Note further that when the data are consistent and $n$ is large, $r^2$ is 
simply given by the variance $\Sigma^2$ of the measurements: 
$r^2w^2 \simeq n \Sigma^2$. The bias is thus by a 
factor $1/(1+n s^2)$, which will become arbitrarily large as the 
number of data points increases. 

The origin of the bias is clear: smaller values of $m_i$ have a smaller 
normalization uncertainty $m_i s$, and are thus preferred in the fit. 
Several examples of situations where this leads to absurd results 
may be found in Ref.~\cite{dagos}. The 
variance of $t$ is afflicted by the same downward bias:
\begin{equation}
\label{eq:dagostinivar}
V_{tt} 
= \frac{\Sigma^2 + s^2 w^2(1+r^2)}{1+ r^2 s^2
  w^2/\Sigma^2}.
\end{equation}
We will henceforth refer to this as the ``d'Agostini bias'', after 
Ref.~\cite{dagos,dagos94} where it was studied and explained.

%Note that when we use the full covariance matrix, the dependence 
%of the $\chi^2$ 
%on the measurements is no longer purely Gaussian, and thus the likelihood 
%function becomes $L(t) = N(t)\exp(-\half\chi^2_m(t))$, where the 
%overall normalization factor $N$ now depends on $t$. 
%Maximizing the likelihood 
%is thus no longer the same as minimizing the $\chi^2$, and thus it should 
%come as no surprise that results dependent on this latter procedure 
%turn out to be biased. In principle it should be possible to correct for 
%this by constructing an error function which includes the normalization 
%factor: we have not attempted to pursue such an approach here.

\subsection{More than one experiment}
\label{subsec:mcovn}

In the second example with $n$ distinct experiments the d'Agostini bias 
is much milder. With the covariance matrix Eq.~(\ref{eq:covmNexp}), 
the $\chi^2$-function is
\begin{equation}
  \label{eq:chi2twoexp}
  \chi_m^2(t) = \sum_{i=1}^n\frac{(t-m_i)^2}{\sigma_i^2+m_i^2s_i^2} 
  \, ,
\end{equation}
which is minimized when 
\be
\label{eq:tnexpm}
t =\frac{ \sum_{i=1}^n \frac{m_i}{\sigma_i^2+s^2m_i^2}}{
\sum_{i=1}^n \frac{1}{\sigma_i^2+s^2m_i^2}}.
\ee

To exhibit the bias, consider again the case when all statistical 
uncertainties are equal, $\sigma_i=\sigma$: then on expanding in powers 
of 
\be\label{rdefeq}
r^2 \equiv \frac{1}{n}\sum_{i=1}^n \frac{(m_i-\bar{m})^2}{\bar{m}^2} = 
\frac{1}{n}\sum_{i=1}^n \frac{m_i^2-\bar{m}^2}{\bar{m}^2},
\ee
with $\bar{m}$ given by Eq.~(\ref{def:mbar}) we find
\begin{equation}
\label{eq:dagostinieq2}
t = 
\bar{m}\left(1-2r^2\frac{s^2\bar{m}^2}{\sigma^2+s^2\bar{m}^2} +O(r^4)\right).
\end{equation}
The origin of the bias is the same as in
Eqs.~(\ref{eq:dagostinieq},\ref{eq:dagostinin}), and indeed for two
data points the biases are  
approximately the same: however for $n$ independent experiments, the 
bias Eq.~(\ref{eq:dagostinieq2}) is stable for large $n$. 
In this case the variance is given by (again expanding in $r$)
\be
\label{eq:dagostinivar2}
V_{tt} = \smallfrac{1}{n}(\sigma^2 + s^2\bar{m}^2(1+  r^2))+O(r^4),
\ee
which is as expected, since $\bar{m}^2(1+r^2) 
= \frac{1}{n}\sum_{i=1}^n m_i^2$.

\section{The Penalty Trick}
\label{sec:ncov}

The standard~\cite{dagos94,dagos,Takeuchi:1995xe} way to include
normalization uncertainties in the Hessian approach while avoiding 
 the d'Agostini bias consists of including
the normalizations of the data $n_i$ as parameters in the fit, with penalty 
terms to fix their estimated value close to one with variance
$s_i^2$. In this section we will discuss this widely used method,
which we  will
refer to as the ``penalty trick'': we will see that 
while it gives correct results for a single experiment, when used to combine 
results from several experiments it is actually still biased. 

\subsection{One Experiment}
\label{subsec:ncov1}

We first consider a single experiment, with covariance matrix 
Eq.~(\ref{eq:covstat}), but now with an 
overall normalization uncertainty with variance $s^2$. The value of $t$ is 
then obtained by minimizing the error function
\be
  \label{eq:chi2norm}
  E_{\rm Hess}(t,n)  
  = \sum_{i=1}^n \frac{(t/n-m_i)^2}{\sigma_i^2}+ \frac{(n-1)^2}{s^2}\, .
\ee
where the last term is called the penalty term. 
%This method will be labelled as the $n-$cov method.
The parameters $t$ and $n$ are then determined by minimizing this error 
function: minimizing with respect to $t$ gives $t=nw$, with $w$ as defined in 
Eq.~(\ref{def:w}), while minimization with respect to $n$ fixes $n=1$, so for
the central values the result is the same as in the Monte Carlo 
approach.\footnote{
Note that to obtain an unbiased result from this approach, the factor 
$n$ must rescale the theory, not the data: if instead of 
Eq.~(\ref{eq:chi2norm}) we took
\be
%  \label{eq:chi2normxxx}
  E_{\rm Hess}  
  = \sum_{i=1}^n\frac{(t-nm_i)^2}{\sigma_i^2} 
+ \frac{(n-1)^2}{s^2}\, .\nonumber
\ee
we would get a result with a strong downward bias similar to that 
in Eq.~(\ref{eq:dagostinin})~\cite{dagos94,dagos}.}

In order to compute the error on the fitted quantity in this approach, 
we need to evaluate the Hessian matrix:
\begin{equation}
  \label{eq:Hessnorm}
  V^{-1} = \half 
    \begin{pmatrix}
    \smallfrac{\partial^2\chi^2}{\partial t^2} &
    \smallfrac{\partial^2\chi^2}{\partial t\partial n} \\
    \smallfrac{\partial^2\chi^2}{\partial n\partial t} &
    \smallfrac{\partial^2\chi^2}{\partial n^2} 
  \end{pmatrix}  = \frac{1}{\Sigma^2}
  \begin{pmatrix}
    1 & -t \\
    -t & \Sigma^2/s^2+t^2
  \end{pmatrix}\, .
\end{equation}
The covariance matrix is obtained by inverting $V^{-1}$. In this way one 
recovers 
\begin{equation}
  \label{eq:Vtt}
  V_{tt}=\Sigma^2+s^2 w^2 \, .
\end{equation}
This is the same result as Eq.~(\ref{eq:varnsim}) obtained with the 
Monte Carlo approach, apart from the cross-correlation 
term between the variances of the measurements and the variances of the 
normalization, akin to the last term in Eq.~(\ref{eq:varnm}). 

For the case of a single 
experiment, we have thus recovered within the
Hessian approach the Monte Carlo result of Sect.~\ref{subsec:Norm}: in
this simple case the Hessian and Monte Carlo methods are (almost) equivalent.
It is straightforward to generalize this 
equivalence to a general covariance matrix Eq.~(\ref{eq:cov}).

\subsection{More than one experiment}
\label{subsec:ncovn}

Let us now turn to the  more complex situation where we have
several data points from different experiments, and thus with 
independent normalizations, $n_i=1\pm s_i$. In this case, the Monte Carlo
result Eq.~(\ref{eq:VartNexp}) is unbiased (in the sense that it 
gives an unbiased average over the data when all uncertainties are equal), 
but it does not satisfy the requirement of decoupling. We will now 
show that the penalty trick leads to a result which does satisfy 
decoupling, but is biased when all uncertainties are equal.

Following the same line of reasoning as for the single experiment 
above, we set up the error function
\begin{equation}
  \label{eq:chi2H2norm}
  E_{\rm Hess}(t,n_i) = \sum_{i=1}^n\frac{(t/n_i-m_i)^2}{\sigma_i^2} +
  \sum_{i=1}^n\frac{(n_i-1)^2}{s_i^2}\,,
\end{equation}
where now we have a separate penalty term for each of the normalizations 
to be fitted. The minimum is obtained for
\begin{eqnarray}
  \label{eq:tmin2norm}
  t &=&\frac{ \sum_{i=1}^n\frac{m_i}{n_i\sigma_i^2}}
         {\sum_{i=1}^n\frac{1}{n_i^2\sigma_i^2}} ,\\
  \label{eq:nmin2norm}
  n_i&=& 1 + \frac{s_i^2 t}{n_i^2\sigma_i^2}
\left(\frac{t}{n_i}-m_i\right)\, .
\end{eqnarray}
These $n+1$ equations are now complicated nonlinear relations which 
must be solved for $t$ and $n_i$. A general analytic solution is 
probably impossible, and it seems very likely that for a large number of 
experiments the number of solutions will grow rapidly, making it 
difficult to select the correct one. However it is possible to find 
solutions for certain special cases, which are sufficient to show that 
the approach is biased.

First we specialize to the case of only two experiments: to explore 
the bias we again assume that $\sigma_1=\sigma_2\equiv\sigma$ and 
$s_1=s_2=s$, as in Sect.~3.1. 
Adding and subtracting the two equations for $n_1$ and $n_2$, and 
substituting the equation for $t$, we find
\begin{equation}
  \label{eq:nplusminus}
  n_1^2+n_2^2 = n_1+n_2\, ,\qquad
  n_1^2-n_2^2= -2\Delta n_1n_2\, ,
\end{equation}
where
\begin{equation}
  \label{eq:Deltadef}
  \Delta = \frac{s^2}{\sigma^2 + m_1m_2s^2}\half (m_1^2-m_2^2)\, .
\end{equation}
Solving these for $n_1$ and $n_2$, and choosing the solution close to one
(note that $\Delta$ will generally be rather small) gives
\begin{eqnarray}
   \label{eq:nsoln}
   n_i &=& \half\left(1 + \frac{1\mp\Delta}{\sqrt{1+\Delta^2}}\right),\\
  \label{eq:tsoln}
    t &=& \bar{m}\frac{\half(1+\sqrt{1+\Delta^2} 
              + r\Delta)}{1+\Delta^2},
\end{eqnarray}
where $\bar{m}$ and $r$ are defined in Eq.~(\ref{def:mrtwo}). 

The variance may now be computed as before by inversion of the three by 
three Hessian matrix. The result is not very enlightening, however it 
simplifies if we expand in powers of $r$: the Hessian method then gives
\begin{eqnarray}
   t &=& \bar{m}\Big(1+ \frac{s^2\bar{m}^2(\sigma^2-2s^2\bar{m}^2)}
           {(\sigma^2+s^2\bar{m}^2)^2}r^2 + O(r^4)\Big),
\label{eq:tHess}\\
  V_{tt} &=& \half\sigma^2 + \half s^2\bar{m}^2\Big(1+r^2 
   -4\frac{(2\sigma^2+s^2\bar{m}^2)^2}
         {(\sigma^2+s^2\bar{m}^2)^3}s^2\bar{m}^2r^2 + O(r^4)\Big) 
\, . \label{eq:tHessvar}
\end{eqnarray}
%\begin{center}
\begin{figure}
\epsfig{width=0.49\textwidth,figure=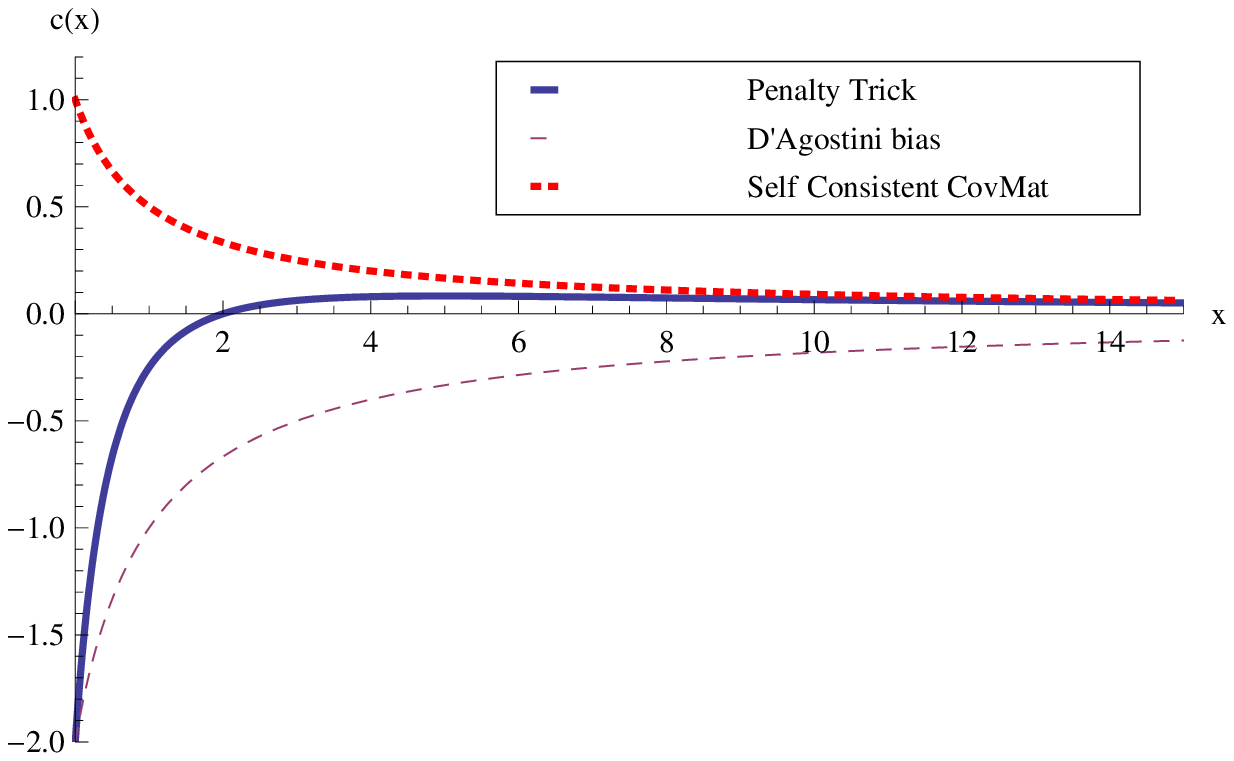}
\epsfig{width=0.49\textwidth,figure=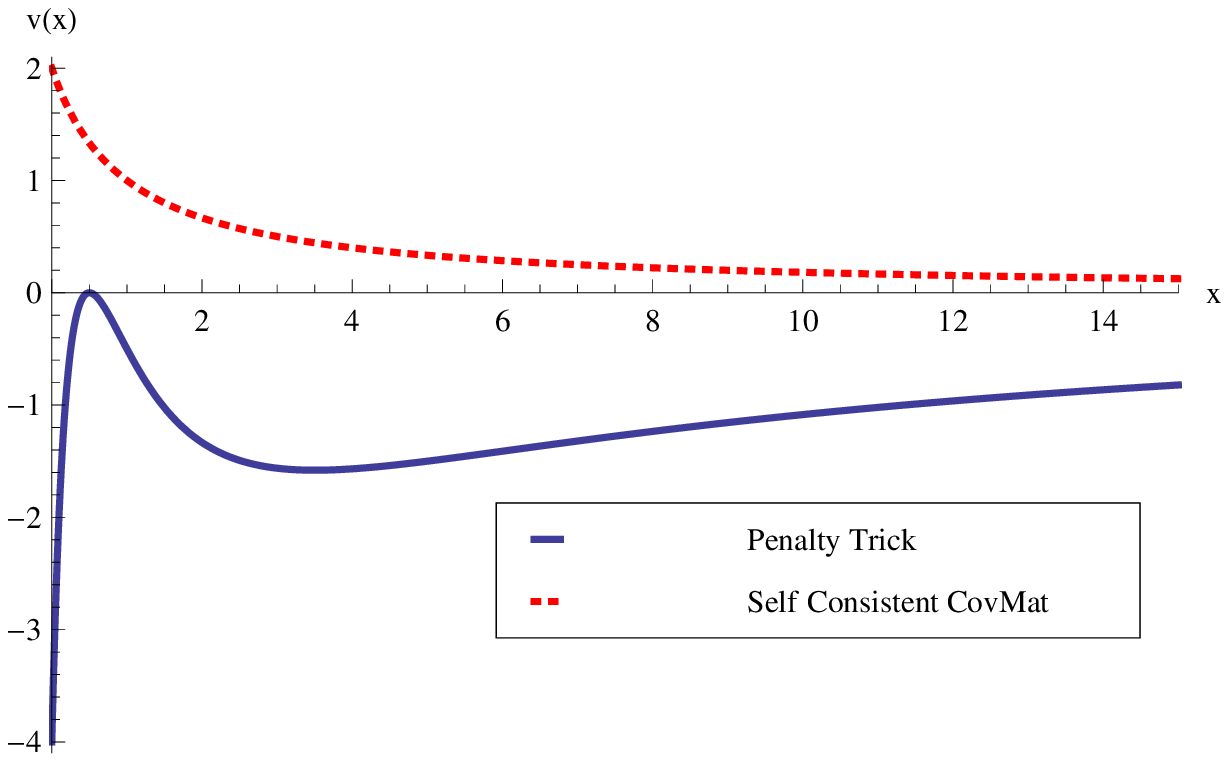}
\caption{\small \label{fig:plot_c_x} The ``bias'' functions $c(x)$
  for the central value (left plot)
  and $v(x)$ for the variance (right plot), as defined in 
  Eqs.~(\ref{eq:tHessx}-\ref{eq:tHessvarx}), corresponding to the
  results obtained when the d'Agostini bias is present,
  Eqs.~(\ref{eq:dagostinieq2}-\ref{eq:dagostinivar2}), using the
  penalty trick, Eqs.~(\ref{eq:tHess}-\ref{eq:tHessvar}), and using
  the self-consistent covariance matrix method,
  Eqs.~(\ref{tsinrexp}-\ref{Vttsinrexp}).  The unbiased result
  corresponds to $c=v=0$. 
%In the case of $v(x)$, the 
%result for the D'Agostini bias $v(x)=0$ is not shown.
}
\end{figure}
%\end{center}

Clearly, both the central value and variance
Eqs.~(\ref{eq:tHess},\ref{eq:tHessvar}) are biased: the normalization
uncertainties leads to data being  weighed
differently based on their central values, even when their
normalization uncertainties are the same. It is not difficult to show 
from the structure of 
Eqs.~(\ref{eq:tmin2norm},\ref{eq:nmin2norm}) that the results 
Eqs.~(\ref{eq:tHess},\ref{eq:tHessvar}) remain true 
for any number of data points, i.e. for all $n\geq 2$, provided 
that $\bar{m}$ and $r$ are defined as in Eqs.~(\ref{def:mbar},\ref{def:r}), 
and an overall factor of $2/n$ is included in the variance.

This bias is more subtle 
than the d'Agostini bias Eq.~(\ref{eq:dagostinieq2}), in that it is caused by  
nonlinearities in the error function rather than by a consistent bias 
in the variances. 
%Again this should come as no surprise: the 
%nonlinearities mean that the likelihood 
%function becomes $L(t) = N(t)\int dn_i\, \exp(-\half E_{\rm Hess}(t,n_i))$, 
%where the normalization factor $N$ again depends on $t$, so again 
%maximizing the likelihood is no longer the same as minimizing the error 
%function. Constructing a maximum likelihood estimator along these lines 
%seems to us a very challenging task, which we have not pursued.
It can thus have either sign, depending on 
the relative weight of statistical and normalization uncertainties: we
can rewrite Eqs.~(\ref{eq:tHess}-\ref{eq:tHessvar}) in the form
\begin{eqnarray}
   t &=& \bar{m}\big(1+ c(\smallfrac{\sigma^2}{s^2\bar{m}^2})r^2 
                         + O(r^4)\big),
\label{eq:tHessx}\\
  V_{tt} &=& \smallfrac{1}{n}\sigma^2 + \smallfrac{1}{n} 
          s^2\bar{m}^2\big(
   1+ r^2 + v(\smallfrac{\sigma^2}{s^2\bar{m}^2})r^2+ O(r^4)\big) 
\, , \label{eq:tHessvarx}
\end{eqnarray}
with the functions $c$ and $v$ given by
\begin{equation}
c(x) = \frac{x-2}{(x+1)^2},\qquad v(x) = -4\frac{(2x-1)^2}{(x+1)^3}.
\label{eq:cvdef}
\end{equation}
 The unbiased results would correspond to $c=v=0$.  
The d'Agostini--biased results
Eqs.~(\ref{eq:dagostinieq2}-\ref{eq:dagostinivar2}) can also be cast in
the form of Eq.~(\ref{eq:tHessx}-\ref{eq:tHessvarx}), but now with
$c(x) = -2/(x+1)$, $v(x)=0$ (since in this case the variance 
is unbiased at $O(r^2)$).

The ``bias'' functions $c(x)$ and $v(x)$ for these two cases, as well
as for a further biased case to be discussed below in
Sect.~\ref{sec:tcov}, are compared in Fig.~\ref{fig:plot_c_x}. Thanks to the
penalty trick, the bias in the
central value is generally less severe, but the variance is now also biased. 
%and moreover the $c$ and $v$ functions now vary rather wildly. 

%------------------------------------------------------------
\begin{figure}
\begin{center}
\epsfig{width=0.75\textwidth,figure=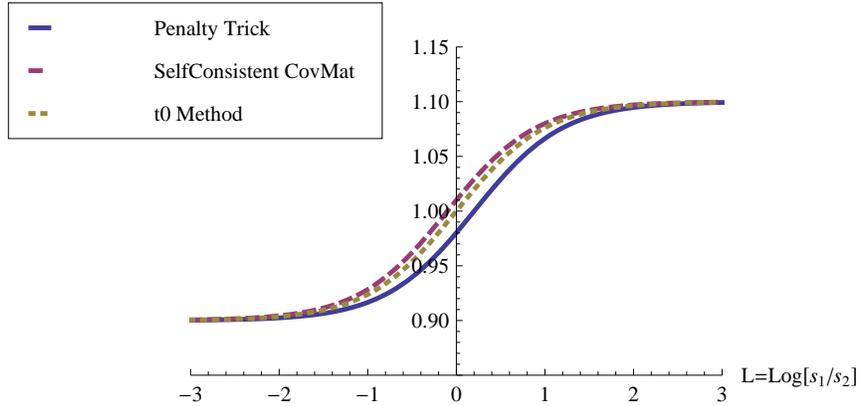}
\end{center}
\caption{\small \label{fig:plot_t}
Dependence of the central value $t$ on the ratio of normalization
uncertainties $s_1$ and $s_2$ for a pair of measurements with
central values $m_1=0.9$, $m_2=1.1$ and negligible uncertainties
$\sigma_i$. The curves correspond to the cases of the penalty trick
method Eq.~(\ref{normdomhessn}), to the self--consistent covariance
matrix method
Eq.~(\ref{tnormdom}), and to the $t_0$ method
Eq.~(\ref{Enormdomz}). The unbiased result must be symmetric about the
point $L=0$ (i.e. $s_1=s_2$): only the $t_0$ curve is 
unbiased. Decoupling of the data point with the larger error is 
clearly visible in the graph for $L\to \pm \infty$.}
\end{figure}
%------------------------------------------------------------

It is also interesting to explore decoupling in this approach. To do this 
we again consider two experiments, in the special case in which the 
normalization errors dominate, 
so $\sigma_1^2,\sigma_2^2\ll s_1^2m_1^2,s_2^2m_2^2$. Then
Eqs.(\ref{eq:tmin2norm},\ref{eq:nmin2norm}) can again be solved: 
$n_i=t/m_i$, with
\be
\label{normdomhess}
t = m_1m_2 \frac{m_1s_1^2 + m_2 s_2^2}
{m_1^2 s_1^2 + m_2^2 s_2^2}.
\ee
The variance is now
\be
\label{normdomhessvar}
V_{tt} = m_1^2m_2^2\frac{s_1^2s_2^2}{m_1^2s_1^2+m_2^2s_2^2}.
\ee
When $s_1\ll s_2$, we recover $t=m_1$ and $V_{tt}=m_1^2s_1^2$, as we should.
However the result Eq.~(\ref{normdomhess}) 
looks rather strange: one would expect 
that in this limit the measurements $m_1$ and $m_2$ to be simply 
weighted by $1/s_1^2$ and $1/s_2^2$, in analogy to the weighting 
Eq.~(\ref{def:w}) in the single experiment case, and the variance to be a 
similar weighted average of $m_1^2$ and $m_2^2$ 
(see Eqs.~(\ref{Enormdomz}--\ref{varnormdomz}) below). Instead the 
weighting is more complicated, reflecting the bias in this method for 
general $s_1\neq s_2$. Note for example that when $s_1=s_2$, 
$t=m_1m_2(m_1+m_2)/(m_1^2+m_2^2)\neq \half(m_1+m_2)$, and 
$V_{tt} = s^2 m_1^2m_2^2/(m_1^2+m_2^2)\neq \smallfrac{1}{4} s^2(m_1^2+m_2^2)$.
The result Eq.~(\ref{normdomhess}) is shown in Fig. \ref{fig:plot_t} for the 
special case $m_1=0.9$ and $m_2=1.1$. 
The bias is readily apparent in the asymmetry of the curve.

The generalization to $n$ independent experiments is straightforward in 
this limit: 
Eq.~(\ref{normdomhess}) becomes
\be
\label{normdomhessn}
t = \frac{\sum_{i=1}^n \frac{1}{m_is_i^2}}{\sum_{i=1}^n \frac{1}{m_i^2s_i^2}},
\ee
so if $s_i=s$, when there should be no bias, we have instead 
$t=\sum_{i=1}^n m_i^{-1}/\sum_{i=1}^n m_i^{-2}\neq \bar{m}$ 
unless $m_i=m$.  

\section{A Self--Consistent Covariance Matrix}

\label{sec:tcov}

We saw in Sect.~\ref{sec:mcov} 
that minimizing the $\chi^2$ Eq.~(\ref{eq:chisqm}) 
constructed using a covariance matrix of the form Eq.~(\ref{eq:covmoneexp})  
gives the so--called d'Agostini
bias. This bias comes from the dependence of the 
normalization term in the covariance matrix on $m_i$ 
and $m_j$: indeed the bias is proportional to the differences
$m_i-m_j$. 
A possible way out, alternative to the penalty trick of the previous
section and  based on a covariance matrix approach,
was suggested by d'Agostini in Ref.~\cite{dagos94}. Namely,
one could  choose to use
\be
\label{covmattil}
(\covt)_{ij}= (\cov)_{ij}+ s^2 t^2 \ ,
\ee
since $t$ is, by construction, a more precise estimator of the observable than
$m_i$, and has already averaged out the differences in central value of 
the different measurements. 
We will now show that this method leads to results which are similar
to those found using the penalty trick: for one experiment there is no bias,
but for several experiments a bias arises. There is also a problem with 
multiple solutions.

\subsection{One experiment}
\label{subsec:tcov1}

With two measurements within a single experiment, we now 
have a covariance matrix
\begin{equation}
\label{covoneexpx}
           (\covt)_{ij}= \left(\begin{array}{cc}\sigma_1^2 + s^2t^2&s^2t^2\cr
                             s^2t^2& \sigma_2^2+s^2t^2\cr\end{array}\right)
\end {equation}
so the $\chi^2$ is
\begin{equation}
  \label{eq:chi2oneexpt}
  \chi^2_t(t) = \frac{(t-m_1)^2(\sigma_2^2+ t^2 s^2) + 
(t-m_2)^2(\sigma_1^2+s^2 t^2) - 2(t-m_1)(t-m_2) s^2 t^2}
{\sigma_1^2\sigma_2^2 + (\sigma_1^2+\sigma_2^2)s^2t^2}
  \, .
\end{equation}
It is easy to check that minimizing this $\chi^2$ with 
respect to $t$ gives again $t=w$, where $w$ is the weighted average 
Eq.~(\ref{def:w}). The variance $V_{tt}$ is now simply 
the inverse of the second derivative of the $\chi^2$ Eq.~(\ref{eq:chi2oneexp}) 
at the minimum: a straightforward but tedious calculation then leads
back to 
Eq.~(\ref{eq:Vtt}).

%Applying this technique in the Monte Carlo method, we now take as the error function
%\begin{equation}
%  \label{eq:Eoneexp}
%  \chi^2_t(T) = \frac{(T-NM_1)^2(\sigma_2^2+ T^2 s^2) + 
%(T-NM_2)^2(\sigma_1^2+s^2 T^2) - 2(T-NM_1)(T-NM_2) s^2 T^2}
%{\sigma_1^2\sigma_2^2 + (\sigma_1^2+\sigma_2^2)s^2T^2}
%  \, .
%\end{equation}
%This error function is then  
%minimized by Eq.\ref{eq:tminnsim}, with the result that 
%$\rm{E}[t]$ and $\rm{Var}[t]$ are again given by Eq.\ref{def:w} 
%and Eq.\ref{eq:varnsim}.

It is not difficult to show that everything works 
out for a general covariance matrix Eq.~(\ref{covmattil}) for a 
single experiment with $n$ data:
\bea
\label{eq:dchi2dtnexp}
\half\frac{\partial\chi^2_t}{\partial t} 
&=& \sum_{i,j=1}^n(\covtinv)_{ij}(t-m_j)
+\half \sum_{i,j=1}^n(t-m_i)\left(\frac{\partial\covtinv}
{\partial t}\right)_{ij}(t-m_j)\nonumber\\
&=& \sum_{i,j=1}^n(\covtinv)_{ij}(t-m_j)
-\half \sum_{i,j,k,l=1}^n(t-m_i)(\covtinv)_{ik}
\left(\frac{\partial\covt}
{\partial t}\right)_{kl}(\covtinv)_{lj}(t-m_j)\nonumber\\
&=& \sum_{i,j=1}^n(\covtinv)_{ij}(t-m_j)
\Big[1- s^2 t \sum_{k,l=1}^n(\covtinv)_{lk}(t-m_k)\Big],
\eea
which vanishes when
\be
\label{eq:tsolnt}
t =\frac{ \sum_{i,j=1}^n(\covtinv)_{ij}m_j}{
\sum_{i,j=1}^n(\covtinv)_{ij}}\, . 
\ee
The variance of $t$ is given by evaluating 
$\frac{\partial^2\chi_t^2}{\partial t^2}$  at the minimum: this yields
\be
\label{eq:vartt}
V_{tt} = \frac{1}{\sum_{i,j=1}^n(\covtinv)_{ij}}\, .
\ee
When 
\be
\label{covmattilmod}
(\covt)_{ij}=
\delta_{ij}\sigma^2_{i}+s^2 t^2 \ ,
\ee
the inverse is
\be
\label{covmattilmodinv}
(\covtinv)_{ij}=
\frac{\delta_{ij}}{\sigma^2_{i}}-\frac{s^2 t^2}{\sigma_i^2\sigma_j^2}
\frac{\Sigma^2}{\Sigma^2+ s^2t^2}  \ ,
\ee
where $\Sigma$ is as defined in Eq.~(\ref{def:Sig}).
Thus Eq.~(\ref{eq:tsolnt}) and Eq.~(\ref{eq:vartt}) simplify to the familiar 
results $t=w$ Eq.~(\ref{def:w}) and $V_{tt} = \Sigma^2+s^2w^2$ 
Eq.~(\ref{eq:Vtt}).

%If we were to use this method in the Monte Carlo approach, we would find instead that
%$t$ is given by Eq.~(\ref{eq:tminnsim}) so ${\mathrm E}[t]= w$, but
%\be
%\label{eq:varnsimt}
%{\rm Var}[t] = \Sigma^2(1+s^2)+ s^2 w^2(1+r^2),
%\ee
%where $r$ is as defined in Eq.~(\ref{def:r}),
%which is a little more precise than Eq.~(\ref{eq:varnsim}) since it correctly 
%includes the fluctuations in the central values.

Note however that since $\chi_t^2$ Eq.~(\ref{eq:chi2oneexp}) is no longer 
quadratic in $t$, there is also a spurious solution: the term in square 
brackets in Eq.~(\ref{eq:dchi2dtnexp}) vanishes when $t=\Sigma^2/ws^2$
which might be troublesome since it may lie close to the correct solution 
$t=w$ whenever $s^2w^2\sim\Sigma^2$.

\subsection{More than one experiment}
\label{subsec:tcovn}

Consider now the case of  $n$ independent experiments. The 
covariance matrix is then
\begin{equation}
\label{covtwoexpx}
          (\covt)_{ij}= (\sigma_i^2+ s_i^2 t^2)\delta_{ij},    
\end {equation}
so the $\chi^2$ is 
\begin{equation}
  \label{eq:chi2twoexpx}
  \chi^2_t(t) = \sum_{i=1}^n\frac{(t-m_i)^2}{\sigma_i^2+s_i^2t^2}  
  \, .
\end{equation}
The minimum of $\chi_t^2$ is found by solving the system of nonlinear equations
\be
\label{eq:nonlinearmess}
 \sum_{i=1}^n
 \frac{(t-m_i)(tm_is_i^2+\sigma_i^2)}{(t^2s_i^2+\sigma_i^2)^2}=0\, . 
\ee
In general these equations will have $4n-2$ solutions, so for a large 
number of experiments finding the correct solution might be difficult.  

To explore the bias we consider as usual the symmetric situation 
$\sigma_i=\sigma$, $s_i=s$, as we did in Sect.~\ref{subsec:ncovn}. 
Then $t$ is the (upper) solution to the quadratic equation
\be
\label{quadratic}
(\bar{m}s^2t + \sigma^2)(t-\bar{m}) = r^2\bar{m}^2s^2 t,
\ee
so again when $r$ is small we have a solution close to $t=\bar{m}$:
\be
\label{tsinrexp}
t=\bar{m}\left(1 + \frac{\bar{m}^2s^2}{\sigma^2+s^2\bar{m}^2} r^2 
+ O(r^4)\right).
\ee
Thus as might be expected from the shape of the $\chi^2$ the 
minimum is pushed upwards for $m_1\neq m_2$. The variance is also biased:
\be
\label{Vttsinrexp}
V_{tt}=\smallfrac{1}{n}\sigma^2+\smallfrac{1}{n}s^2\bar{m}^2
\Big(1+r^2 + \frac{2s^2\bar{m}^2}
{\sigma^2+s^2\bar{m}^2}r^2 + O(r^4)\Big).
\ee
The biases can again be cast in the
form of Eq.~(\ref{eq:tHessx}) and Eq.~(\ref{eq:tHessvarx}) but now
with
\be
\label{eq:sconstcv}
c(x) = \frac{1}{x+1},\quad v(x)=\frac{2}{x+1}.
\ee 
They are compared to those of the penalty trick method in  
Fig.~\ref{fig:plot_c_x}.

Another interesting special case is when the normalization errors 
are dominant, so (for $n$ experiments)
$s_im_i \gg \sigma_i$. The solution for $t$ then reduces to
\be
\label{tnormdom}
t = \frac{\sum_{i=1}^n \frac{m_i^2}{s_i^2}}{\sum_{i=1}^n \frac{m_i}{s_i^2}}.
\ee
So when one of the $s_i$ is very large, this experiment 
decouples as expected. However 
the weighting is still biased: when $s_i=s$, $t = 
\sum_i m_i^2/\sum_i m_i \neq \bar{m}$ unless all $m_i=m$. The result 
Eq.~(\ref{tnormdom}) is also plotted in Fig. \ref{fig:plot_t}: it is 
instructive to compare it with the superficially similar 
result Eq.~(\ref{normdomhessn}) obtained with the penalty trick, which was 
also biased, but in the opposite direction. 
%The variance is now
%\be
%\label{varnormdom}
%V_{tt} = \frac{1}{\sum_{i=1}^2 1/s_i^2}
%\frac{(\sum_{i=1}^n m_i^2/s_i^2)^2}{(\sum_{i=1}^n m_i/s_i^2)^2} 
%\ee
%to be compared with Eq.\ref{normdomhessvar}. 

Thus the self--consistent covariance matrix discussed here is 
biased when 
used for more than one experiment, like the penalty trick method of
Sect.~\ref{sec:ncov}. Since these methods are biased in
the Hessian approach, they would also fail to provide
a suitably unbiased error function for fitting to  
Monte Carlo replicas. Moreover these methods have multiple 
solutions, which also make them difficult to implement in a 
Monte Carlo, since if some of the replicas are fitted to the 
wrong solution, these will clearly also lead to an incorrect final result. 
In the next section, we will present
a new method which is free of multiple solutions, is unbiased 
when uncertainties are equal, but which unlike the method of 
Sect.~\ref{subsec:Norm} 
also correctly weights the different 
experiments according to their normalization uncertainties when these 
are unequal. 

\section{Unbiased Fitting}
\label{sec:t0cov}

The biases and multiple solutions found in the previous section come 
from the fact that the 
$\chi^2(t)$ function used in the fitting is no longer a quadratic function 
of the observable $t$ being fitted, and thus the distribution of 
$\exp(-\half\chi_t^2(t))$ is no longer Gaussian. The dependence of the 
covariances of the data on $t^2$ distorts the shape of the $\chi^2$, 
and thus introduces a bias. The only way to avoid this is to hold the 
covariance matrix fixed when performing the fitting. We can do this by 
evaluating the covariance matrix using some fixed value $t_0$ rather 
than $t$. The value of $t_0$ can then be tuned independently to be 
consistent with the value of $t$ obtained from the fit. The basic idea
is then to determine $t_0$ self--consistently in an iterative 
way. The use of theoretical estimates 
to avoid d'Agostini bias through an iterative procedure can also be found 
in the treatment of multiplicative systematic errors by the 
H1 Collaboration \cite{HERA}.

%In the language of likelihood, the likelihood 
%function now becomes $L(t) = N_{t_0}\exp(-\half \chi^2_{t_0}(t))$.  
%Maximizing the likelihood is thus the same as minimizing $\chi^2_{t_0}(t)$ 
%with respect to $t$, holding $t_0$ fixed, 
%and thus yields (asymptotically) unbiased estimators.

The necessity for such a procedure is particularly clear in the Monte Carlo 
approach. When we fit to replicas, we do so on the assumption that each 
replica is generated with a given distribution of the data central 
values according to their experimental uncertainties. Clearly this 
uncertainty should be fixed once and for all, not 
vary from replica to replica.  We will now show that this procedure
indeed gives unbiased results for equal normalization uncertainties 
and that it also satisfies the decoupling
criterion, for both the Hessian and Monte Carlo methods. Finally we also 
show that the iterative determination of $t_0$ converges very rapidly, and 
thus that the method is also practical.

\subsection{One experiment}
\label{subsec:t0cov1}

For a single experiment, in place of Eq.~(\ref{covmattil})
the covariance matrix is chosen to be
\be
\label{covmathat}
(\covtz)_{ij}= (\cov)_{ij}+t_0^2 s^2,
\ee
where $t_0$ should be viewed as a guess for $t$, to be fixed beforehand. 
In a Hessian approach the $\chi^2$ is then 
\be
\label{chicovhat}
\chi_{t_0}^2(t)
=\sum_{i,j=1}^n
(t-m_i)(\covtzinv)_{ij}(t-m_j),
\ee
and minimization is trivial:
\be
\label{eq:tsolnt0}
t = \frac{\sum_{i,j=1}^n(\covtzinv)_{ij}m_j}{\sum_{i,j=1}^n(\covtzinv)_{ij}},
\ee
while
\be
\label{eq:Vttsolnt0}
V_{tt} = \frac{1}{\sum_{i,j=1}^n(\covtzinv)_{ij}}.
\ee
For the special case in which the data have 
uncorrelated statistical errors only, Eq.~(\ref{eq:covstat}), using the 
inversion Eq.~(\ref{covmattilmodinv}) (with $t$ replaced by $t_0$) we
thus recover $t=w$ Eq.~(\ref{def:w}), independent of the value 
chosen for $t_0$, while 
\be
\label{eq:Vttz}
V_{tt} = \Sigma^2 + s^2 t_0^2.
\ee
This reduces to Eq.~(\ref{eq:Vtt}), but only if we tune $t_0=t$. 

In a Monte Carlo approach the $\chi^2$ for each replica is  
\be
\label{chicovhatmc}
\chi_{t_0}^2(T)
=\sum_{i,j=1}^n
(T-NM_i)(\covtzinv)_{ij}(T-NM_j),
\ee
and minimization is again trivial:
\be
\label{eq:tsolnt0mc}
t = \frac{\sum_{i,j=1}^n(\covtzinv)_{ij}NM_j}{\sum_{i,j=1}^n(\covtzinv)_{ij}}. 
\ee
The expectation value and variance of $t$ can now be straightforwardly 
evaluated: 
\bea
\label{eq:Ett0}
{\rm E}[t] &=& \frac{\sum_{ij=1}^n(\covtzinv)_{ij}m_j}{\sum_{i,j=1}^n(\covtzinv)_{ij}},\\ 
\label{eq:Vartt0}
{\rm Var}[t] &=& 
(1+s^2)\left(\frac{1}{\sum_{i,j=1}^n (\covtzinv)_{ij}}-s^2t_0^2\right)
+ s^2 {\mathrm E}[t]^2.
\eea
This coincides with the Hessian result
Eqs.~(\ref{eq:tsolnt0}-\ref{eq:Vttsolnt0}) when $t_0={\mathrm E}[t]$. Also 
it reduces again to the old results ${\rm E}[t]=w$ Eq.~(\ref{def:w}) and 
${\rm Var}[t]=(1+s^2)\Sigma^2+w^2$ Eq.~(\ref{eq:varnsim}) 
for uncorrelated statistical errors 
Eq.~(\ref{eq:covstat}). These results are then quite independent of 
the value of $t_0$, explaining the success of the old error function 
Eq.~(\ref{eq:efNone}). Note however that the value of the $\chi^2$ 
Eq.~(\ref{chicovhat}) at the minimum for each replica will still 
depend on $t_0$, and will indeed be quite different from that of 
the error function Eq.~(\ref{eq:efNone}), to which it reduces 
only when $t_0=0$.

\subsection{More than one experiment}
\label{subsec:t0covn}

When we have $n$ independent experiments (each with one data point), we 
now choose in place of Eq.~(\ref{covtwoexpx})
\begin{equation}
\label{covtwoexpz}
          (\covt)_{ij}= (\sigma_i^2+ s_i^2 t_0^2)\delta_{ij},    
\end {equation}
so the $\chi^2$ is now
\begin{equation}
  \label{eq:chi2twoexpz}
  \chi^2 = \sum_{i=1}^n\frac{(t-m_i)^2}{\sigma_i^2+s_i^2t_0^2} 
  \, ,
\end{equation}
whence we have the single solution 
\be
\label{eq:tnexpz}
t = \frac{\sum_{i=1}^n \frac{m_i}{\sigma_i^2+s_i^2t_0^2}}{\sum_{i=1}^n \frac{1}{\sigma_i^2+s_i^2t_0^2}},
\ee
and
\be
\label{eq:Vttznexp}
V_{tt} = \frac{1}{\sum_{i=1}^n \frac{1}{\sigma_i^2+s_i^2t_0^2}}.
\ee

For the special case $s_i=s$, $\sigma_i=\sigma$, these reduce to 
$t=\bar{m}$ and $V_{tt} = \frac{1}{n}(\sigma^2+s^2t_0^2)$ 
as they should: the 
fit is unbiased, and the variance is correctly estimated provided only that 
$t_0=t$. When instead the normalization 
uncertainties dominate, 
\be\label{nexpnormdomtz}
t=\frac{\sum_{i=1}^n \frac{m_i}{s_i^2}}{\sum_{i=1}^n \frac{1}{s_i^2}},\qquad
V_{tt}=\frac{t_0^2}{\sum_{i=1}^n \frac{1}{s_i^2}},
\ee 
and thus the central value exhibits decoupling (if some $s_i$ is 
much larger than the others, $t$ becomes independent of the corresponding 
measurement $m_i$), and the variance is again correctly estimated whenever 
$t_0=t$.

In the Monte Carlo approach the $\chi^2$ for a given replica is now
\begin{equation}
  \label{eq:chi2twoexpzmc}
  \chi^2 = \sum_{i=1}^n\frac{(t-N_iM_i)^2}{\sigma_i^2+s_i^2t_0^2} 
  \, .
\end{equation}
Minimization is again straightforward: 
\be
\label{eq:tnexpzmc}
t = \frac{\sum_{i=1}^n \frac{N_iM_i}{\sigma_i^2+s_i^2t_0^2}}{\sum_{i=1}^n \frac{1}{\sigma_i^2+s_i^2t_0^2}},
\ee
whence on averaging over replicas
\bea
\label{eq:nexpEtz}
{\rm E}[t] &=& \frac{\sum_{i=1}^n \frac{m_i}{\sigma_i^2+s_i^2t_0^2}}{\sum_{i=1}^n \frac{1}{\sigma_i^2+s_i^2t_0^2}},\\ 
\label{eq:nexpVartz}
{\rm Var}[t] &=& 
\frac{\sum_{i=1}^n \frac{\sigma_i^2+s_i^2(m_i^2+\sigma_i^2)}
{(\sigma_i^2+s_i^2t_0^2)^2}}{\left(
\sum_{i=1}^n \frac{1}{\sigma_i^2+s_i^2t_0^2}\right)^2}. 
\eea
The central value coincides with the Hessian result
Eq.~(\ref{eq:tnexpz}), while the variance is now determined more accurately: 
Eq.(\ref{eq:nexpVartz}) reduces to Eq.~(\ref{eq:Vttznexp}) whenever 
$m_i^2+\sigma_i^2\sim t_0^2$. 

In the special cases considered previously, in the symmetric 
case $s_i=s$, $\sigma_i=\sigma$ we have 
\be
\label{eq:unbiasedx}
{\rm E}[t]=\bar{m},\qquad 
{\rm Var}[t] = \smallfrac{1}{n}(\sigma^2(1+s^2) + s^2\bar{m}^2(1+r^2)),
\ee 
which is indeed unbiased. Note that unlike in 
the Hessian method, the variance cross term is now properly included (see 
Eq.~(\ref{eq:varnm})), the spread of values $m_i$ now contributes 
to the variance, as it should (see Eq.~(\ref{eq:dagostinivar2})), 
and both the expectation value 
and variance are actually independent of $t_0$, just as they were for the 
case of a single experiment. 

When the normalization uncertainties dominate, $\sigma_i^2 \ll s_i^2t_0^2$
\bea
\label{Enormdomz}
{\rm E}[t]&=&\frac{\sum_{i=1}^n \frac{m_i}{s_i^2}}
{\sum_{i=1}^n \frac{1}{s_i^2}},\\
\label{varnormdomz}
 {\rm Var}[t]&=&\frac{\sum_{i=1}^n \frac{m^2_i}{s_i^2}}
{\left(\sum_{i=1}^n \frac{1}{s_i^2}\right)^2}.
\eea 
The result Eq.~(\ref{Enormdomz}) is compared in Fig.~\ref{fig:plot_t}
to the biased results Eq.~(\ref{tnormdom}) and 
Eq.~(\ref{normdomhess}) obtained previously. It is manifestly
unbiased. Furthermore, 
when the normalization uncertainty of one of the experiments is 
particularly large, this experiment now decouples from both the mean and 
the variance just as it should. Note again that the results 
Eq.~(\ref{Enormdomz}) and Eq.~(\ref{varnormdomz}) are entirely 
independent of $t_0$: in the Monte Carlo method $t_0$ only 
controls the relative balance between the statistical and normalization 
errors, and then only when these are different amongst themselves.

\subsection{Determining $t_0$}
\label{subsec:t0fix}

The results of Sect.~\ref{subsec:t0cov1}-\ref{subsec:t0covn} imply that 
the $t_0$-covariance matrix Eq.~(\ref{covmathat}) gives a $\chi^2$ 
function that can be used to give unbiased fits to the individual
replicas, with $t_0$ controlling
the relative balance between statistical and normalization
errors, both in the Hessian and Monte Carlo  method.
The remaining difficulty with this approach is that $t_0$ is 
not determined self--consistently within the minimization, but rather 
must be fixed beforehand. Clearly if the value chosen is incorrect, this 
may itself lead to an incorrect fit. Indeed, we have
seen that the Monte Carlo and Hessian results with the $t_0$--method lead 
to the same central prediction only when $t_0={\mathrm E}[t]$.

However, the dependence on $t_0$ is rather weak.
That this is the case is qualitatively clear:
firstly $t_0$ only determines the uncertainties, so an error we make in $t_0$ 
is a second order effect; furthermore all dependence on
$t_0$ cancels when all the $m_i$ are equal, when $\sigma_i$ and $s_i$ are 
equal, or when normalization errors dominate over statistical, or 
indeed vice versa. 
To make this more precise, consider a small shift 
$t_0\to t_0+\delta t_0$. The corresponding shift in ${\rm E}[t]$ 
Eq.~(\ref{eq:nexpEtz}) (or its Hessian counterpart Eq.~(\ref{eq:tnexpz}) 
is then given by
\be
\label{eq:delnexpEtz}
\delta{\rm E}[t] = \delta t_0 
\frac{\sum_{i,j=1}^n t_0\frac{(m_i-m_j)(s_i^2\sigma_j^2-\sigma_i^2s_j^2)}
{(\sigma_i^2+s_i^2t_0^2)^2(\sigma_j^2+s_j^2t_0^2)^2}}{
\left(\sum_{i=1}^n \frac{1}{\sigma_i^2+s_i^2t_0^2}\right)^2}.
\ee
As expected this vanishes when all the $m_i$ are equal, 
in the symmetric case $s_i=s$, $\sigma_i=\sigma$, and 
in the limits when either statistical or normalization uncertainties 
dominate. Elsewhere, we expect it to be very small. 

To quantify this, 
we first note that when combining a large number of experiments, so 
that $n$ in Eq.~(\ref{eq:delnexpEtz}) is large, the result is 
essentially independent 
of $n$ (since both numerator and denominator grow as $n^2$), so a 
typical contribution to the sums may be taken as indicative of the 
overall result. Now for any pair of measurements, 
$2t_0(m_i-m_j)\sim m_i^2-m_j^2$ will typically be of the same size as 
the uncertainties in the measurements, so if all percentage uncertainties 
are of the same order $\Delta$, i.e. $s_i\sim \Delta$, 
$\sigma_i\sim \Delta t_0$, then Eq.~(\ref{eq:delnexpEtz}) 
gives $\delta{\rm E}[t]\sim \Delta^2 \delta t_0 $. So since $\Delta$ is 
always rather less than one, we always expect $\delta{\rm E}[t]\ll\delta t_0$.
 
It follows that $t_0$ can be determined iteratively: a first determination
of $t$ is performed with a zeroth--order guess for $t_0$, such as, for
example, $t_0=0$, which in the Monte Carlo approach corresponds 
to the simple choice
Eq.~(\ref{eq:efNoneNexp}) in which normalization uncertainties are not
included in the error function. The result for ${\rm E}[t]$ thus obtained is
used as $t_0$ for a second iteration, and so on. Since 
$\delta{\rm E}[t]\sim \Delta^2 \delta t_0$, and $\Delta \ll 1$, it is clear 
that these iterations will converge rapidly to a result within the envelope 
of the purely statistical uncertainties on ${\rm E}[t]$. Moreover, 
since as we saw in the previous section ${\rm E}[t]$ is unbiased for any 
$t_0$, the final result (which has $t_0 = {\rm E}[t]$) will also be 
unbiased. Note incidentally that this also implies that the 
$t_0$-method must give a  result different in general to that 
obtained with the penalty trick, since the latter is biased (see
Fig.2).
%\footnote{An iterated version of the
%penalty trick method has been also been
%suggested~\cite{HERA}: whereas it is unclear to us whether this avoids
%the bias of the penalty trick method, it is similar in spirit to the
%$t_0$ method developed here}.

The estimate Eq.~(\ref{eq:delnexpEtz}) then  suggests that already the first
iteration should be sufficient to provide a determination with a
relative uncertainty of $\Delta^2$ for data affected by typical
relative uncertainties $\Delta$, i.e. the procedure is expected
to converge at the first iteration for all practical purposes.

In table~\ref{tab:covmat_summary} we summarise
the features of our new  $t_0$--method compared to those of all 
the other procedures discussed in this paper, based on the two criteria of
freedom from bias when uncertainties are all equal, and decoupling when 
the normalization uncertainty of one experiment is very large, as
presented in the end of
Sect.~\ref{subsec:Hess}. Specifically, we consider the Monte Carlo method with
normalization uncertainties not included in the error function
discussed in Sect.~\ref{subsec:Norm}; the d'Agostini--biased Hessian
method of Sect.~\ref{sec:mcov}; the Hessian method with penalty trick
presented in Sect.~\ref{sec:ncov}; the self--consistent covariance
matrix method ($t$-cov) discussed in
Sect.~\ref{sec:tcov}, and finally the $t_0$--method discussed in this
section.
The two central columns in the table give the ratio of the central
value to the unbiased results Eq.~(\ref{eq:Ett0}) 
and Eq.~(\ref{eq:nexpEtz}) in the
one--experiment and $n$--experiment cases respectively, thereby
exposing the bias whenever the ratio is not equal to one. The result 
is given for uncorrelated systematic uncertainties, all equal to $\sigma$,
and normalization uncertainties all equal to $s$, and assuming that the 
difference between measurements $r$ Eq.~(\ref{rdefeq}) is
$r^2\approx \sigma^2/\bar{m}^2$. The last
column gives the central value for a fit to two experiments, in the 
limit where 
normalization uncertainties are dominant and other uncertainties can
be neglected, in order to show how (or if) an experiment decouples when 
its normalization uncertainty becomes very large.
\begin{table}
\begin{center}
\scriptsize
\begin{tabular}{|c|c|c|c|c|}
\hline
method & ${\mathrm E}$ or $\chi^2$& \multicolumn{2}{|c|}{Bias for $s_i=s$, $\sigma_i=\sigma$} & 
Decoupling $s_i^2m_i^2 \gg \sigma_i^2$   \\
\hline
\multicolumn{2}{|c|}{} & 1 expt, $n$ data & $n$ expts & 
 2 expts \\
\hline
\hline
MC with $t_0=0$& (\ref{eq:efNone}),(\ref{eq:efNoneNexp})  & $1$ & $1$ 
& $\frac{1}{2}\lp m_1+m_2\rp$  \\
\hline
Hessian with d'Agostini bias &  (\ref{eq:chi2oneexp}),(\ref{eq:chi2twoexp}) &$\frac{1}{1+ns^2}$  &   
$1-\frac{2s^2\sigma^2}{\sigma^2+s^2\bar{m}^2}+\ldots$   
& $\frac{1/m_1s_1^2 + 1/m_2s_2^2}
{1/m_1^2s_1^2 + 1/m_2^2s_2^2}$ \\ 
\hline
Hessian with penalty trick & (\ref{eq:chi2norm}),(\ref{eq:chi2H2norm}) & $1$ 
& $1-\frac{s^2(\sigma^2-2s^2\bar{m}^2)}{(\sigma^2+s^2\bar{m}^2)^2}+\ldots$ &
$\frac{1/m_1s_1^2 + 1/m_2s_2^2}
{1/m_1^2s_1^2 + 1/m_2^2s_2^2}$ \\ 
\hline
$t$-cov. mat.  & (\ref{eq:chi2oneexpt}),(\ref{eq:chi2twoexpx})& $1$ & 
$1+\frac{s^2\sigma^2}{\sigma^2+s^2\bar{m}^2}+\ldots$ & 
$\frac{m_1^2/s_1^2 + m_2^2/s_2^2}{m_1/s_1^2+m_2/s_2^2}$  \\
\hline
$t_0$-cov. mat.  
&(\ref{chicovhatmc}),(\ref{eq:chi2twoexpzmc}) & $1$ & $1$ 
& $\frac{m_1/s_1^2 + m_2/s_2^2}{1/s_1^2+1/s_2^2}$ \\
\hline
\end{tabular}
\end{center}
\caption{\small \label{tab_covmat_summary} Summary of the various
methods for the inclusion of normalization uncertainties. The second
column provides the reference to the pair of $E$-- or $\chi^2$--functions whose
minimization determines the results for  one experiment and
many  experiments in each case. Results for the bias (defined in the text) 
are given assuming the data asymmetry $r^2\approx \sigma^2/\bar{m}^2$.}
\label{tab:covmat_summary}
\end{table}

\subsection{Likelihood}

So far in this paper we have taken a naive approach, in which we constructed
least squares estimators, and minimised them 
with respect to the theoretical prediction $t$. It is interesting 
to consider instead how one might construct a likelihood function for 
the measurements and normalizations, and thus whether any 
of our estimators are maximum likelihood estimators.

Consider for definiteness the case of several experiments, with measurements 
$m_i$ and variances $\sigma_i$. In presenting the measurements in this 
form, there is an underlying assumption that the measurements are Gaussian.
The likelihood is then simply defined as the probability that the measurements
take the observed values given a certain theoretical value $t$ for such 
measurements:
\be
\label{L0}
P(m|t) = N \exp\Big( -\half \sum_{i=1}^n \frac{(m_i-t)^2}{\sigma_i^2}\Big),
\ee
where $N$ is some overall normalisation factor for the probability, 
dependent on 
$\sigma_i$ but not on $m_i$ or $t$.  Of course this probability is just 
$N\exp(-\half\chi^2(t))$, so the maximum likelihood 
estimator is found by minimising the $\chi^2$, i.e. it is the same 
as the least squares estimator. 

Now consider what happens when there are also normalization uncertainties 
$n_i$ with variances $s_i$. Again in the absence of further information 
it is natural to assume these are also Gaussian. Furthermore they are clearly 
entirely independent of the measurement uncertainties, since the physics 
involved in determining the normalization is generally quite independent 
of that related to the measurements $m_i$ or indeed the theoretical value $t$.
Thus the total likelihood should factorise: $P(m,n|t)=P(m|t)P(n)$. The 
maximum likelihood estimators obtained from $P(m,n|t)$ and $P(m|t)$ should 
thus be the same.

Now the Hessian method of Sec.~\ref{sec:mcov}, by 
adopting the $\chi^2$-function Eq.(\ref{eq:chi2twoexp}), assumes for 
the likelihood  
\be
\label{Lm}
 P_m(m|t) = N \exp\Big( -\half \sum_{i=1}^n 
\frac{(m_i-t)^2}{\sigma_i^2+s_i^2m_i^2}\Big).
\ee
This is incorrect, because it is no longer Gaussian in $m_i$, and indeed
not even properly normalised (to normalise it, $N$ must depend on $t$, and 
then maximising $P_m(m|t)$ is no longer the same as minimising the $\chi^2$).
Thus the $\chi^2$-function Eq.(\ref{eq:chi2twoexp}) is not a maximum 
likelihood estimator, principally because the probability
distribution it assumes is skewed by the normalization uncertainties.

Similarly the Hessian method with penalty trick
presented in Sec.~\ref{sec:ncov}, by adopting the error 
function Eq.(\ref{eq:chi2H2norm}), assumes for the likelihood 
\be
\label{Ln}
 P_n(m,n|t) = N \exp\Big( -\half \sum_{i=1}^n 
\Big[\frac{(m_i-t/n_i)^2}{\sigma_i^2} +\frac{(n_i-1)^2}{s_i^2}\Big]\Big).
\ee
This is also incorrect, because now the likelihood, while Gaussian in $m_i$, 
is not Gaussian in $n_i$, and furthermore cannot be factorised into a 
product $P(m|t)P(n)$. Once again it is not properly normalised: $N$ must 
depend on $t$. So this too does not give us a maximum likelihood estimator.
The main problem here is that since the model for the likelihood does not 
factorise, the assumption of a common theoretical result $t$ introduces 
artificial correlations between 
the normalization measurements $n_i$; this leads to biases since these 
measurements are in principle completely independent. This is 
why the penalty trick, while giving correct results for a single experiment 
with only one overall normalization uncertainty, fails when applied to 
several independent experiments.

The self--consistent covariance matrix method ($t$-cov) discussed in
Sec.~\ref{sec:tcov}, with $\chi^2$-function Eq.(\ref{eq:chi2twoexpx}),
assumes the likelihood to be
\be        
\label{Lt} 
 P_t(m|t) = N \exp\Big( -\half \sum_{i=1}^n \frac{(m_i-t)^2}
{\sigma_i^2 + t^2s_i^2}\Big).
\ee
This is better, because it is now Gaussian in $m_i$. However the 
normalization $N$ still depends on $t$, and 
thus the maximum likelihood estimator no longer minimises the $\chi^2$, but has
a correction from $\ln N(t)$. It may thus be possible to construct the 
maximum likelihood estimator by this route, but it is difficult because
everything is so nonlinear in $t$.

Finally consider the $t_0$--method discussed in this
section, which takes as its starting point the $\chi^2$-function 
Eq.(\ref{eq:chi2twoexpz}). Here the assumption for the likelihood is
\be
\label{Lt0}
             P_{t_0}(m|t) = N \exp\Big( -\half \sum_{i=1}^n 
\frac{(m_i-t)^2}{\sigma_i^2 + t_0^2s_i^2}\Big)
\ee 
This is now correct: it is Gaussian in $m_i$, properly normalised, and all
we have to do to make sure we get the correct answer is choose $t_0$ 
consistently. Note that it can alternatively be formulated as
\be  
\label{Lt0n}
  P_{t_0}(m,n|t) = N \exp\Big( -\half \sum_{i=1}^n 
\Big[\frac{(m_i-t/n_0)^2}{\sigma_i^2 + t_0^2s_i^2} 
+ \frac{(n_i-n_0)^2}{s_i^2}\Big]\Big).
\ee 
This is rather like the penalty trick Eq.(\ref{Ln}), but with the $n_i$ 
in the first term replaced
with its best estimate $n_0$, just as in going from $m$-cov Eq.(\ref{Lm}) 
to $t_0$-cov Eq.(\ref{Lt0}) we
replace $m_i$ in the denominator with our best estimate $t_0$. Note of course
that actually $n_0=1$: this is the natural choice for presenting the data
that all experimentalists choose. Eq.(\ref{Lt0n}) is also a good 
definition of the 
likelihood: it is Gaussian in both $m_i$ and $n_i$, 
the normalization is 
determined quite independently of $t$, and it factorises correctly into 
the product $P_{t_0}(m|t)P(n)$. Thus when we minimise wrt $t$, 
$P_{t_0}(m|t)$ and $P_{t_0}(m,n|t)$ give the same maximum likelihood 
estimator for $t$, as they should.

Since the $t_0$-method yields the maximum likelihood estimator, it possesses 
all the nice asymptotic properties of maximum likelihood estimators: in 
particular it is consistent and unbiased, and asymptotically efficient (see
for example Ref.~\cite{Eadie}). 

\section{Application to PDF determination}
\label{sec:apppdf}

 {
 \begin{table}
 \scriptsize
 \centering
 \begin{tabular}{|c|c|c|c|c|c|}
\hline
\multicolumn{6}{|c|}{{\bf DIS data} (NNPDF1.2~\cite{NNPDF12} data set)}\\
  \hline
\hline
  Experiment & Set & STAT (\%)& 
   SYS (\%) & NORM (\%)& TOT (\%) \\ \hline
 NMC-pd   &  & & & &\\ \hline
 &  NMC-pd   &   2.0 &   0.4 &   0.0 &   2.1 \\
 \hline
 NMC      &  & & & &\\ \hline
 &  NMC      &   3.7 &   2.3 &   2.0 &   5.0 \\
 \hline
 SLAC     &  & & & &\\ \hline
 &  SLACp    &   2.7 &   0.0 &   2.2 &   3.6 \\
 \hline
 &  SLACd    &   2.5 &   0.0 &   1.8 &   3.1 \\
 \hline
 BCDMS    &  & & & &\\ \hline
 &  BCDMSp   &   3.2 &   2.0 &   3.2 &   5.5 \\
 \hline
 &  BCDMSd   &   4.5 &   2.3 &   3.2 &   6.6 \\
 \hline
 ZEUS     &  & & & &\\ \hline
 &  Z97lowQ2 &   2.5 &   3.0 &   2.1 &   4.7 \\
 \hline
 &  Z97NC    &   6.2 &   3.1 &   2.1 &   7.8 \\
 \hline
 &  Z97CC    &  33.6 &   5.6 &   2.0 &  34.2 \\
 \hline
 &  Z02NC    &  12.2 &   2.1 &   1.8 &  12.7 \\
 \hline
 &  Z02CC    &  38.6 &   6.2 &   1.8 &  39.3 \\
 \hline
 &  Z03NC    &   6.9 &   3.2 &   2.0 &   8.3 \\
 \hline
 &  Z03CC    &  29.1 &   5.6 &   2.0 &  29.8 \\
 \hline
 H1       &  & & & &\\ \hline
 &  H197mb   &   2.8 &   2.0 &   1.7 &   3.9 \\
 \hline
 &  H197lwQ2 &   2.7 &   2.5 &   1.7 &   4.2 \\
 \hline
 &  H197NC   &  12.5 &   3.2 &   1.5 &  13.3 \\
 \hline
 &  H197CC   &  27.5 &   4.6 &   1.5 &  28.1 \\
 \hline
 &  H199NC   &  14.7 &   2.8 &   1.8 &  15.2 \\
 \hline
 &  H199CC   &  25.5 &   3.8 &   1.8 &  25.9 \\
 \hline
 &  H199NChy &   7.2 &   1.7 &   1.8 &   7.7 \\
 \hline
 &  H100NC   &   9.4 &   3.2 &   1.5 &  10.4 \\
 \hline
 &  H100CC   &  20.4 &   3.8 &   1.5 &  20.9 \\
 \hline
 CHORUS   &  & & & &\\ \hline
 &  CHORUSnu &   4.2 &   6.4 &   7.9 &  11.2 \\
 \hline
 &  CHORUSnb &  13.8 &   7.8 &   8.7 &  18.7 \\
 \hline
 FLH108   &  & & & &\\ \hline
 &  FLH108   &  47.2 &  53.3 &   5.0 &  71.9 \\
 \hline
 NTVDMN   &  & & & &\\ \hline
 &  NTVnuDMN &  16.2 &   0.0 &   2.1 &  16.3 \\
 \hline
 &  NTVnbDMN &  26.6 &   0.0 &   2.1 &  26.7 \\
 \hline
 ZEUS-H2  &  & & & &\\ \hline
 &  Z06NC    &   3.8 &   3.7 &   2.6 &   6.4 \\
 \hline
 &  Z06CC    &  25.5 &  14.3 &   2.6 &  31.9 \\
 \hline
 \end{tabular}
 \caption{\small Different sources of uncertainty for 
  data included in the NNPDF1.2~\cite{NNPDF12}
analysis, which is representative of the
deep-inelastic scattering data included
in typical parton fits. All uncertainties are given
  as a percentages, obtained as averages over all points of 
the percentage values
  in the corresponding set; ``stat'' denotes all uncorrelated uncertainties
  (mostly statistical but also uncorrelated systematics); ``sys''
  uncertainties which are correlated between all experiments in a
  set; ``norm''  multiplicative (normalization) uncertainties which
  are also correlated between all experiments in a
  set; and ``tot'' the average of the sum in quadrature of all
  uncertainties. Normalization uncertainties are 
fully correlated between the NMCp and NMCd datasets and 
cancel in the NMC-pd ratio. \label{tab:unc1}}
\end{table}
}

{
\begin{table}
\scriptsize
\centering
\begin{tabular}{|c|c|c|c|c|c|}
\hline
\multicolumn{6}{|c|}{{\bf  Hadronic data} (NNPDF2.0~\cite{NNPDF20} data set)}\\
  \hline
\hline
 Experiment & Set & STAT (\%)& 
   SYS (\%) & NORM (\%)& TOT (\%) \\
 \hline
 DYE605   &  & & & &\\ \hline
 &  DYE605   &  16.6 &   0.0 &  15.0 &  22.6 \\
 \hline
 DYE886   &  & & & &\\ \hline
 &  DYE886p  &  20.4 &   0.0 &   6.5 &  22.1 \\
 \hline
 &  DYE886d  &  18.3 &   0.0 &   6.5 &  20.6 \\
 \hline
 &  DYE886r  &   3.6 &   1.0 &   0.0 &   3.8 \\
 \hline
 CDFWASY  &  & & & &\\ \hline
 &  CDFWASY  &   4.2 &   4.2 &   0.0 &   6.0 \\
 \hline
 CDFZRAP  &  & & & &\\ \hline
 &  CDFZRAP  &   5.1 &   6.0 &   6.0 &  11.5 \\
 \hline
 D0ZRAP   &  & & & &\\ \hline
 &  D0ZRAP   &   7.6 &   0.0 &   6.1 &  10.2 \\
 \hline
 CDFR2KT  &  & & & &\\ \hline
 &  CDFR2KT  &   4.5 &  21.1 &   5.8 &  23.0 \\
 \hline
 D0R2CON  &  & & & &\\ \hline
 &  D0R2CON  &   4.4 &  14.3 &   6.1 &  16.8 \\
 \hline
 \end{tabular}
\caption{\small Same as Table~\ref{tab:unc1}
but for the hadronic data which is included, in addition to
the DIS data of Table~\ref{tab:unc1}, in the 
NNPDF2.0~\cite{NNPDF20} analysis. Again, this dataset
is representative of the
hadronic data  included
in typical parton fits. Normalization uncertainties are 
fully correlated between the DYE886p and DYE886d datasets and 
cancel in the DYE866r ratio. \label{tab:unc2}}
 \end{table}
}

As mentioned in the introduction, in PDF determinations 
an accurate result is sought while combining
many different
sources of uncertainty for a large number of experiments. Typical
uncertainties in a global parton fit are summarized in
Table~\ref{tab:unc1}  for deep--inelastic scattering (DIS) 
experiments and Table~\ref{tab:unc2}
for hadronic experiments. Specifically,
we list in Tables~\ref{tab:unc1} and~\ref{tab:unc2} the uncertainties of
the DIS data included in the
NNPDF1.2~\cite{NNPDF12} analysis, and those of the Drell-Yan,
weak boson production and jet data
included in the  
NNPDF2.0~\cite{NNPDF20} global fit. The total number
of data points included in these sets is about 3000 for the NNPDF1.2
analysis, and about 3500 for NNPDF2.0. The datasets used of other
recent global parton fits~\cite{MSTW,CTEQ} are similar. For DIS
data, all
uncertainties including normalizations are of the order of a few percent,
while for hadronic data uncertainties vary widely and can be as
large as 20\%.
We expect therefore the impact of a full treatment of
normalization uncertainties to be moderate when fitting DIS data, and
more dramatic in truly global fits which include hadronic data.

\subsection{Impact of normalization uncertainties}
\label{sec:pdf}

Normalization uncertainties have been included in recent
PDF determinations
either by using the penalty trick with the Hessian method in recent
MSTW~\cite{MSTW} and CTEQ~\cite{CTEQ,nadolsky} fits, or
using the Monte Carlo method with the 
$t_0=0$ error function Eq.~(\ref{eq:efNoneNexp}) in recent NNPDF
fits\footnote{In fact, the
  error functions used by NNPDF differs from 
  Eq.~(\ref{eq:efNoneNexp}) by a  factor 
  of $N_i^2$ in the denominator. It is not difficult to see that the only 
  effect of this extra factor is to introduce a small downward bias 
  of relative order $s_i^2$ (so $\sim0.01\%$ for the data in NNPDF1.2) 
  in ${\rm E}[t]$ and ${\rm Var}[t]$.}~\cite{NNPDF10,NNPDF12}. 

As we have seen, neither of these  procedures is entirely
satisfactory. However, the bias in the penalty trick method used by 
MSTW and CTEQ, with uncertainties such as those 
of Tables~\ref{tab:unc1} and~\ref{tab:unc2} is 
likely to be at or below the percent level for DIS data, though 
it could be non-negligible for some hadronic (in particular Drell-Yan) 
data. Similarly, the error in using
the Monte Carlo method with $t_0=0$ can be estimated using
Eq.~(\ref{eq:delnexpEtz}) which for DIS experiments
gives $\delta{\rm E}[t]\sim
0.01\,\delta t_0$. Hence, the NNPDF1.0 and NNPDF1.2 fits performed in 
Ref.~\cite{NNPDF10,NNPDF12} to DIS data are expected
to be within a few percent of the true result.
The deviation is more significant for hadronic data.

\subsection{Implementation of the $t_0$ method}
\label{sec:implement}

%------------------------------------------------------------
\begin{figure}
\begin{center}
\epsfig{width=0.45\textwidth,figure=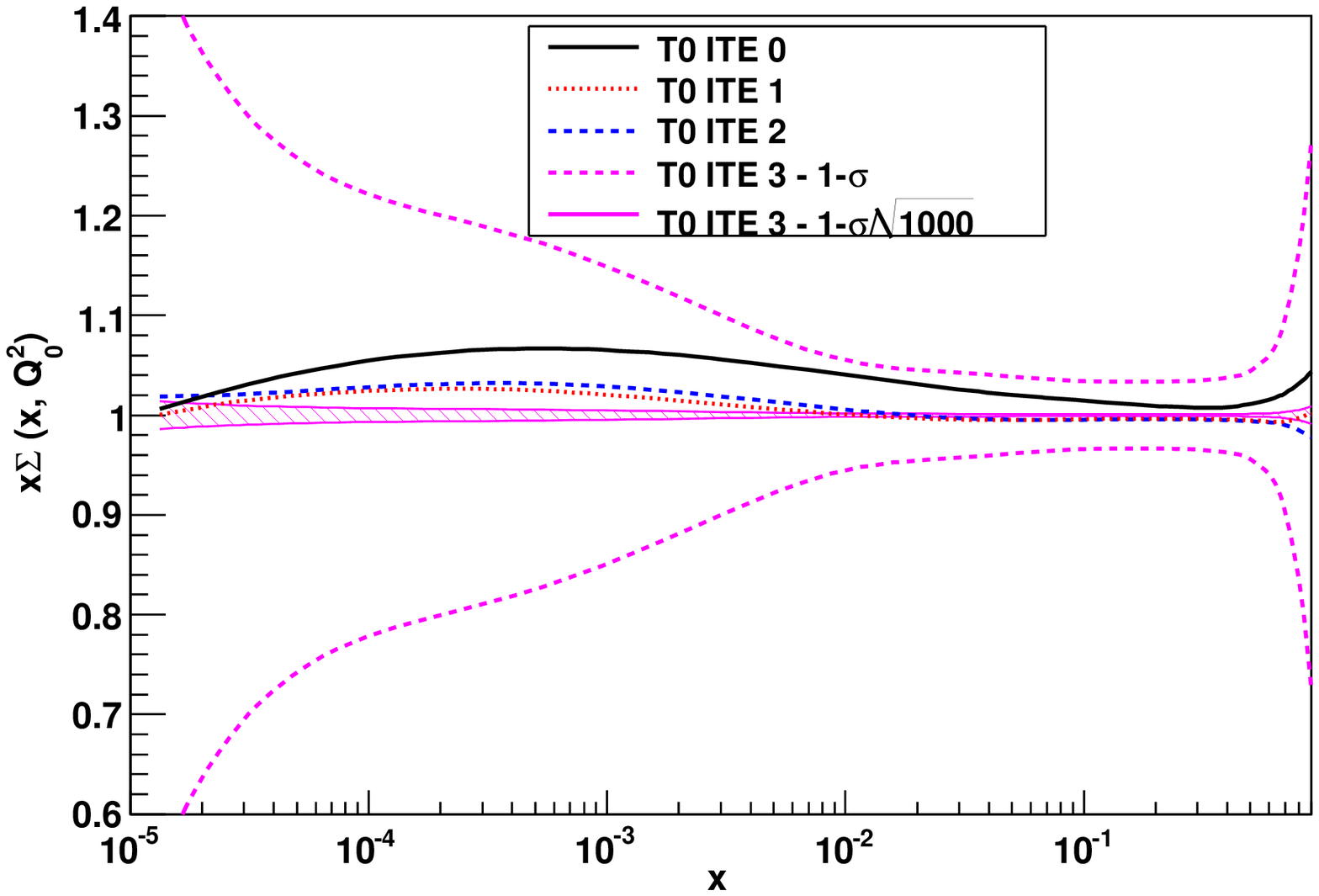}
\epsfig{width=0.45\textwidth,figure=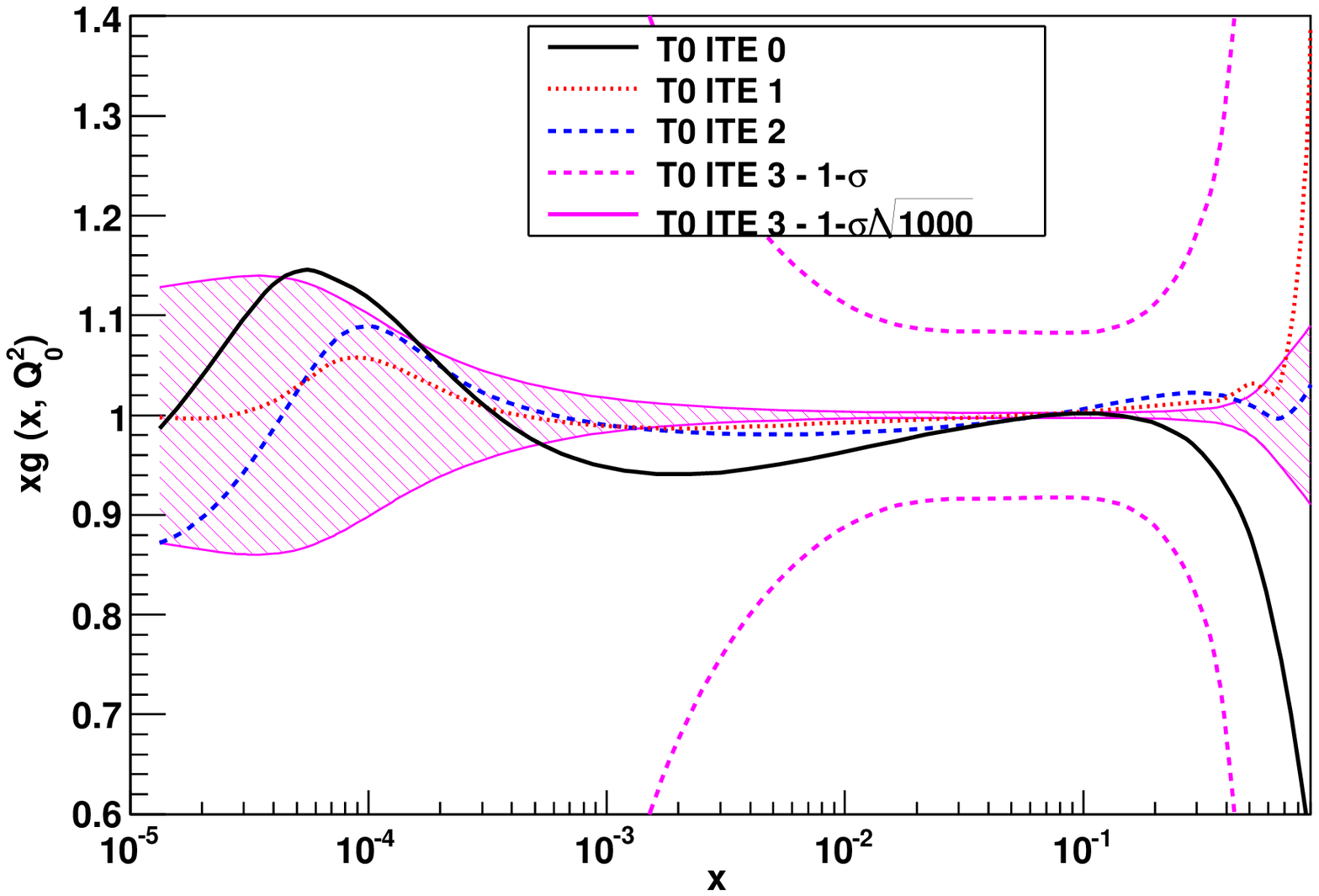}
\epsfig{width=0.45\textwidth,figure=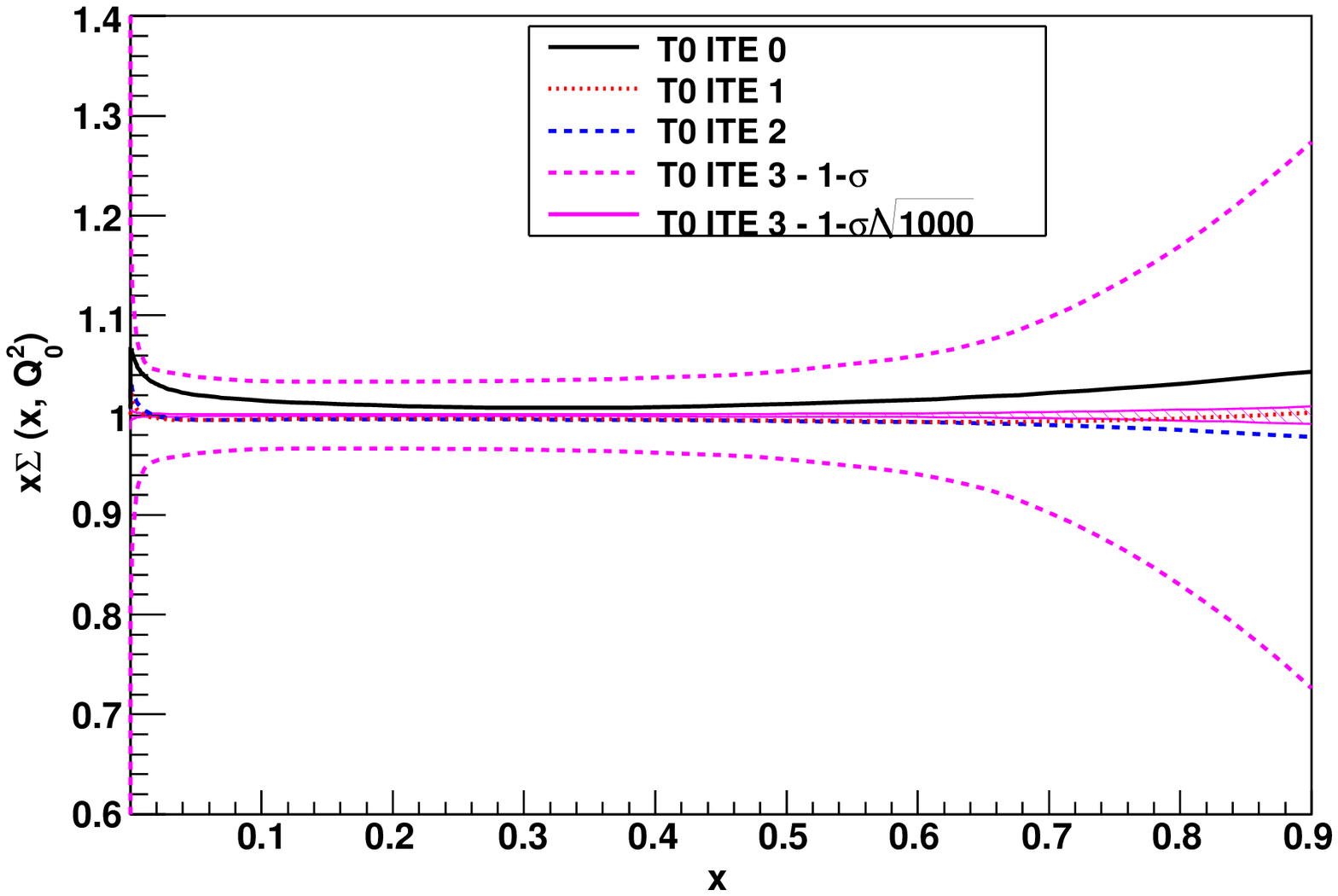}
\epsfig{width=0.45\textwidth,figure=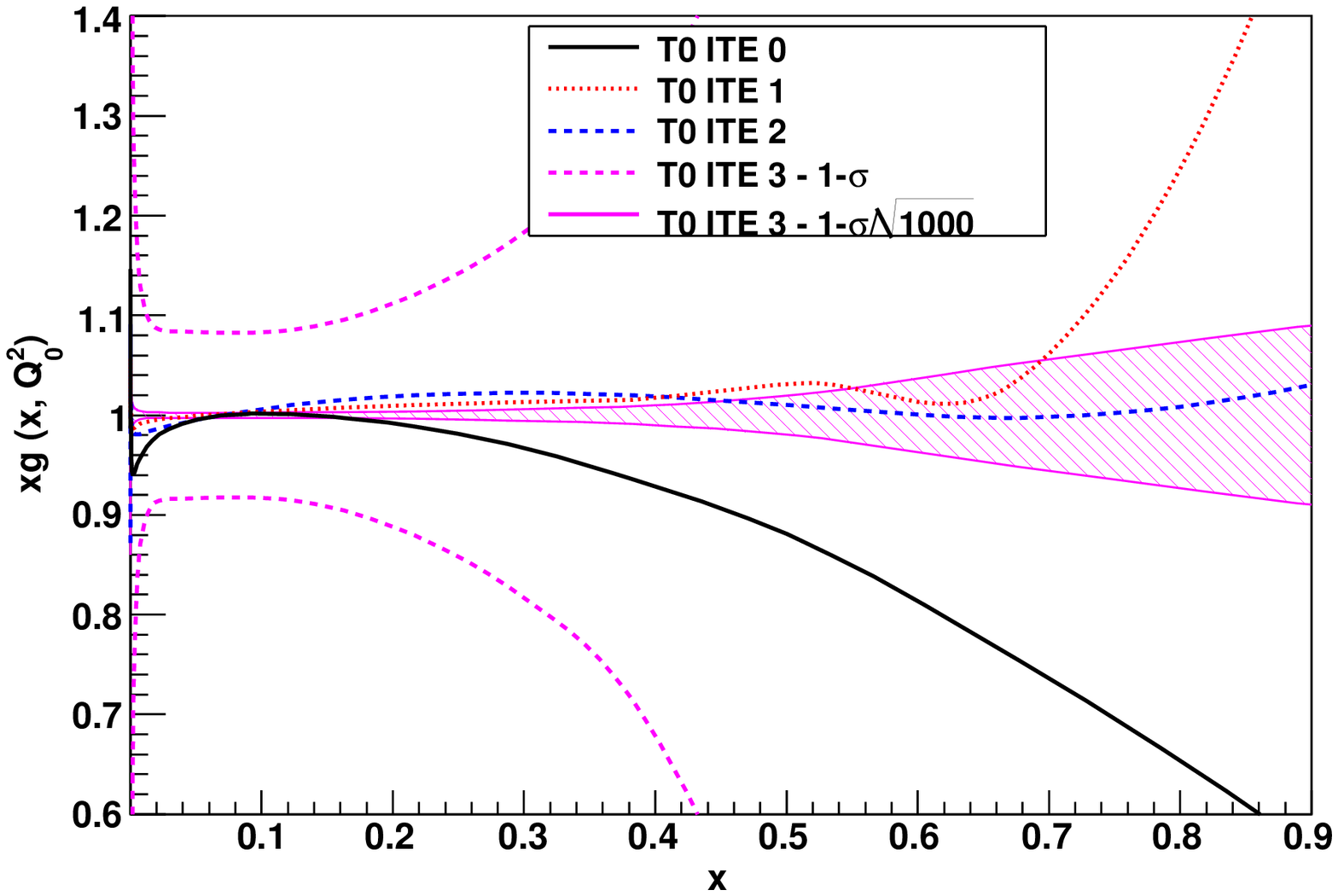}
\epsfig{width=0.45\textwidth,figure=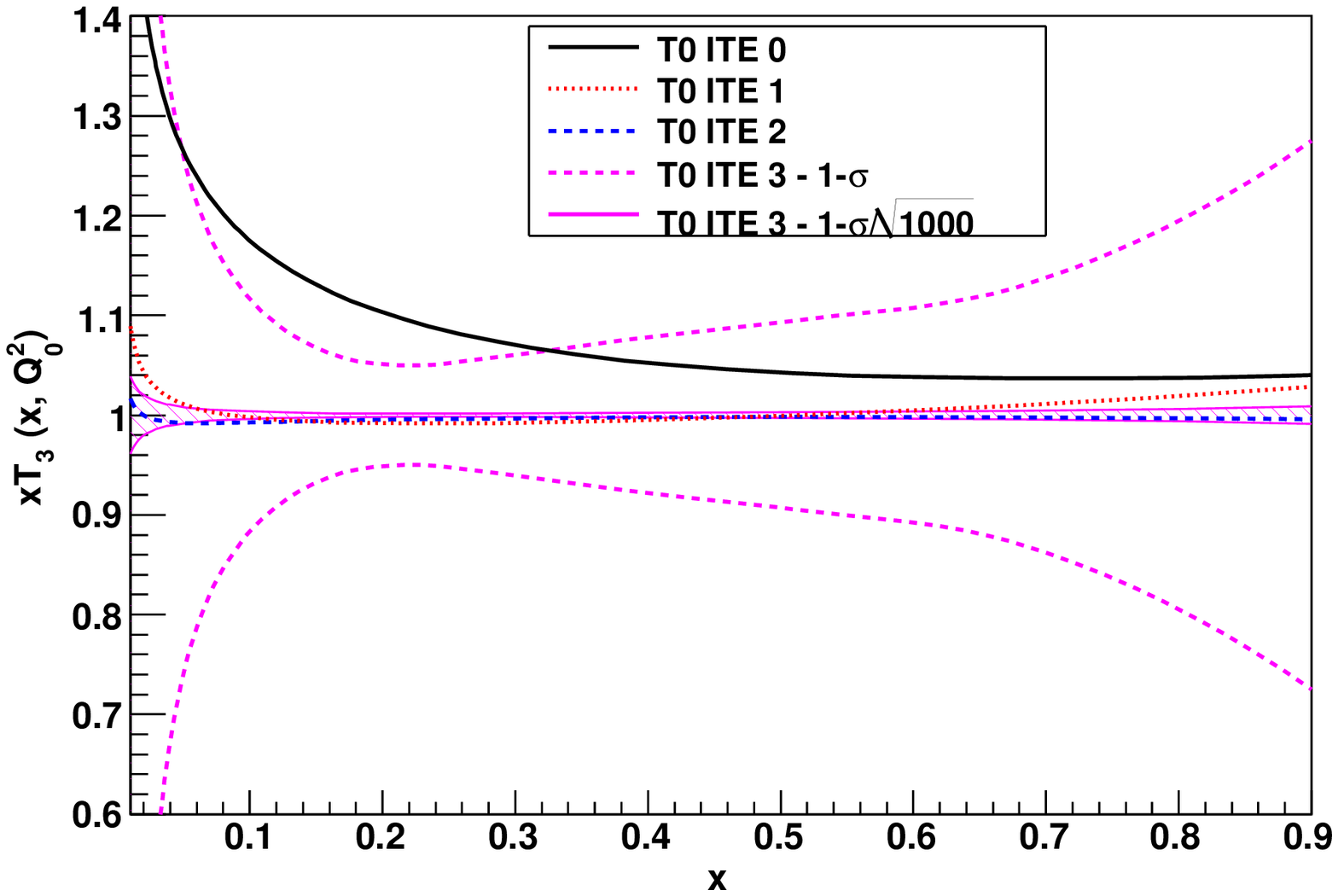}
\epsfig{width=0.45\textwidth,figure=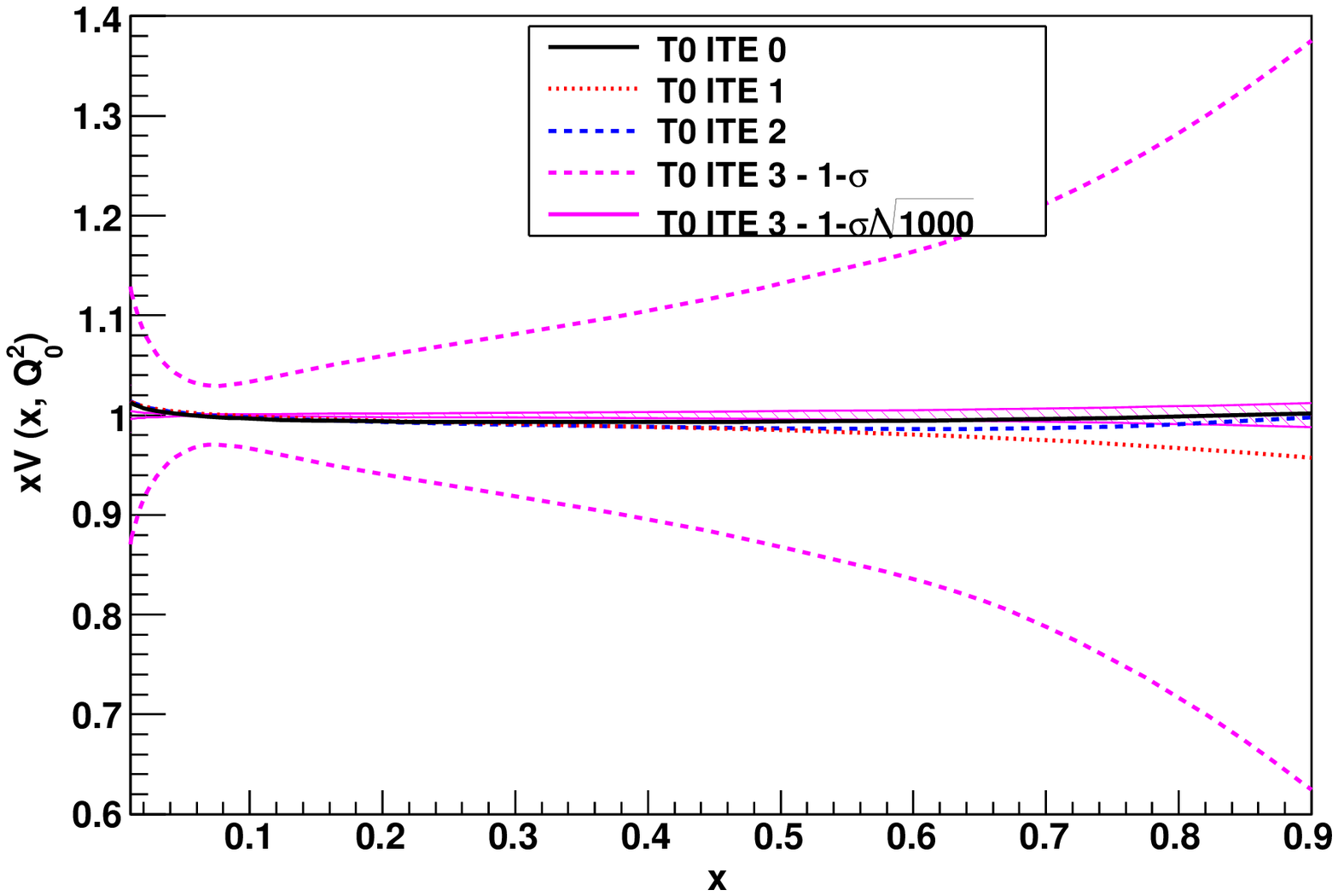}
\epsfig{width=0.45\textwidth,figure=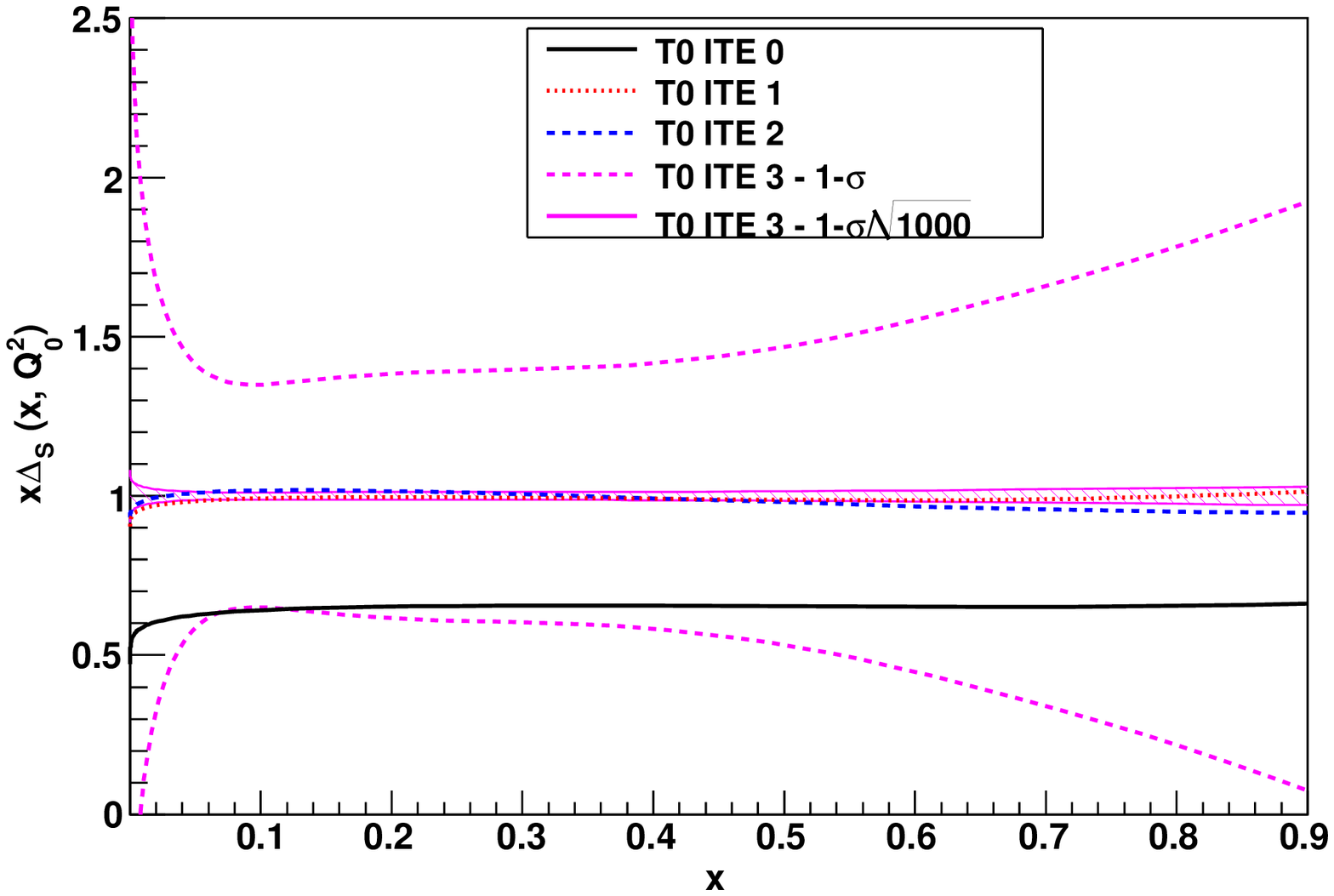}
\epsfig{width=0.45\textwidth,figure=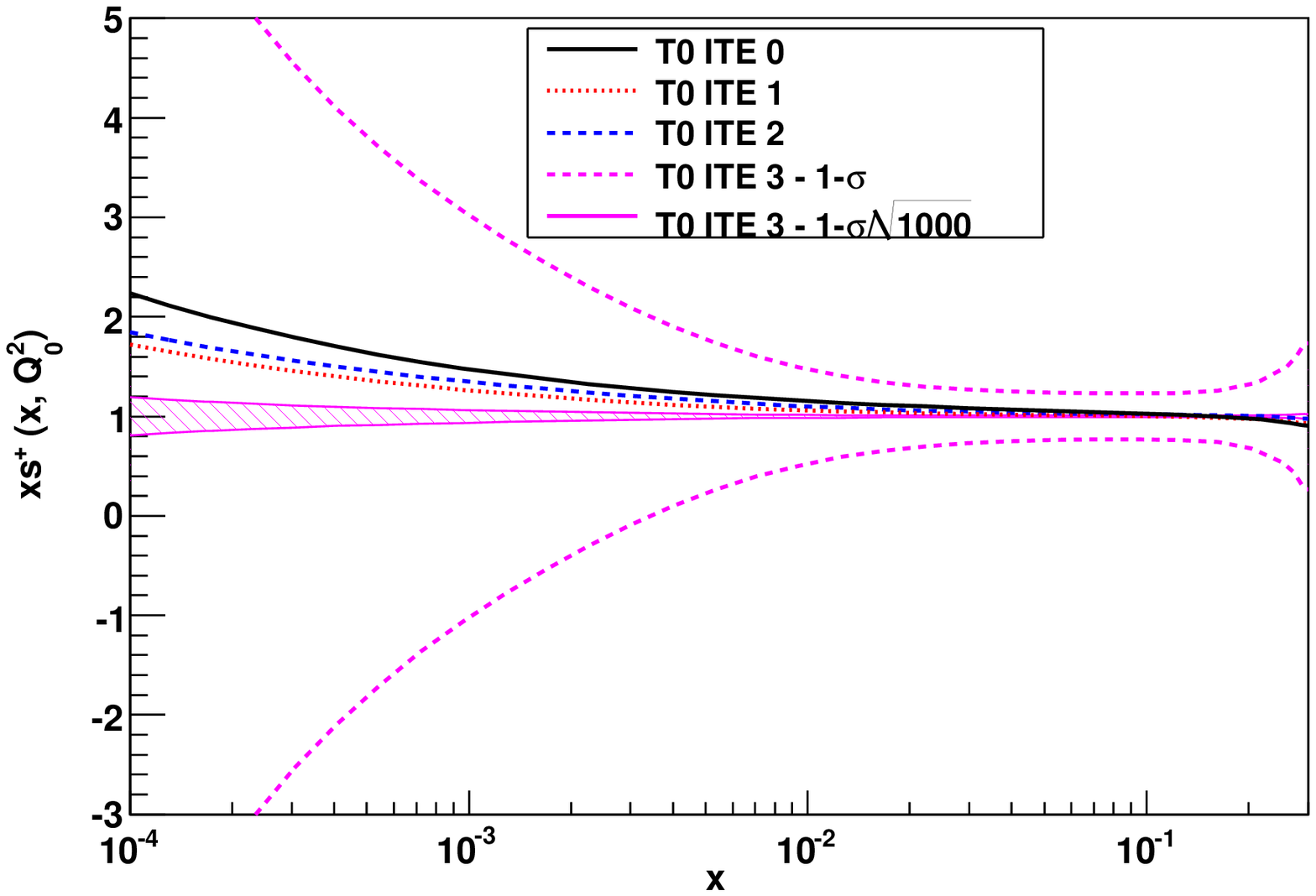}
\end{center}
\caption{\small \label{fig:pdfplots2-1000}
Comparison between NNPDF1.2 PDFs and the results of
the three first $t_0$--iterations. In each case the PDFs shown
have been used to compute the $t_0$ covariance matrix for the next iteration.
Each iterations has been computed with $1000$ Monte Carlo replicas.
We show the one-sigma PDF uncertainty band for
ite3, as well as the same band divided by $\sqrt{1000}$,
which corresponds to the expected statistical fluctuations of the 
PDF central values (hatched area). For the singlet and the
gluon PDFs plots displaying both the small $x$ and large $x$ regions 
are shown.}
\end{figure}
%------------------------------------------------------------
In order to test our general arguments, and also
as a practical illustration of the new
$t_0$ method we have implemented it in the NNPDF framework and
repeated the NNPDF1.2 fit~\cite{NNPDF12} with the $t_0$ method.
As a starting point we take the old NNPDF1.2 fit, and thus $t_0=0$. 
We then performed several iterations in each one
taking the central value of the previous fit to determine $t_0$, and thus 
the covariance matrix Eq.~(\ref{covmathat}):
specifically ite1 thus takes $t_0$ to be the central value of NNPDF1.2
(referred to hereafter as ite0), ite2 takes $t_0$ to be the central 
value of the ite1 fit, and so on.
In this way we can assess the convergence of the method. Three iterations 
proves to be more than sufficient.
At each iteration we produce one thousand replicas: the uncertainty on $t_0$
is then the overall pdf uncertainty divided by $\sqrt{1000}$, which is 
sufficiently small that random fluctuations are kept under control.
% ------------------------------------------------
\begin{table}
\begin{center}
{\footnotesize
\begin{tabular}{|c|c|c|c|}
\hline 
&  {\bf ite1} & {\bf ite2} & 
{\bf ite3}\\
\hline 
$\Sigma(x,Q_0^2)$  & 9.2 & 2.6 & 2.8 \\
\hline
$g(x,Q_0^2)$ & 2.9 & 1.9 & 2.1\\
\hline
 $T_3(x,Q_0^2)$& 7.5 & 0.9 & 0.7 \\
\hline
 $V(x,Q_0^2)$  & 1.1 & 0.8 & 2.8 \\ 
\hline
 $\Delta_S(x,Q_0^2)$  & 14.4 & 1.6 & 1.2 \\
\hline
 $s_+(x,Q_0^2)$  & 2.8 &  3.5 & 3.3 \\
\hline
 $s_-(x,Q_0^2)$  &   1.9 & 1.2 & 2.6 \\ 
\hline
\end{tabular}
}
\end{center}
\caption{\small The stability distances for
various PDFs for the various
iterations of the  NNPDF1.2 $t_0$ fits,
all computed from $1000$ replicas. For each
fit, distances are computed with respect
to the fit from the previous column
(ite1 being with respect to NNPDF1.2). The distances are averaged
over the data regions, as defined for
each PDF in~\cite{NNPDF12}. \label{tab:stabtab2}}
\end{table}
% ---------------------------------------------

In Fig.~\ref{fig:pdfplots2-1000} we show the results for the central pdfs
(i.e. $t_0$) in the various iterations, ite0, ite1, ite2 and ite3 
normalized  with respect to ite3. We show, besides the overall
one-sigma PDF uncertainty of ite3 (the very broad band in the plots), also 
the expected range for the fluctuations of the $t_0$ iterations 
in the convergence regime (the relatively narrow band). As expected, 
convergence is reached very quickly, essentially at the first iteration: 
while the original central value (ite0) often lies some distance from the
central narrow band, all subsequent iterations lie more or less within it. 
Note however that even ite0 lies essentially 
within the broad PDF uncertainty band: the 
shift in central values is generally within one sigma of the 
overall uncertainty.
The only exception is the triplet $T_3$ in the valence region.

\begin{figure}
\begin{center}
\epsfig{width=0.49\textwidth,figure=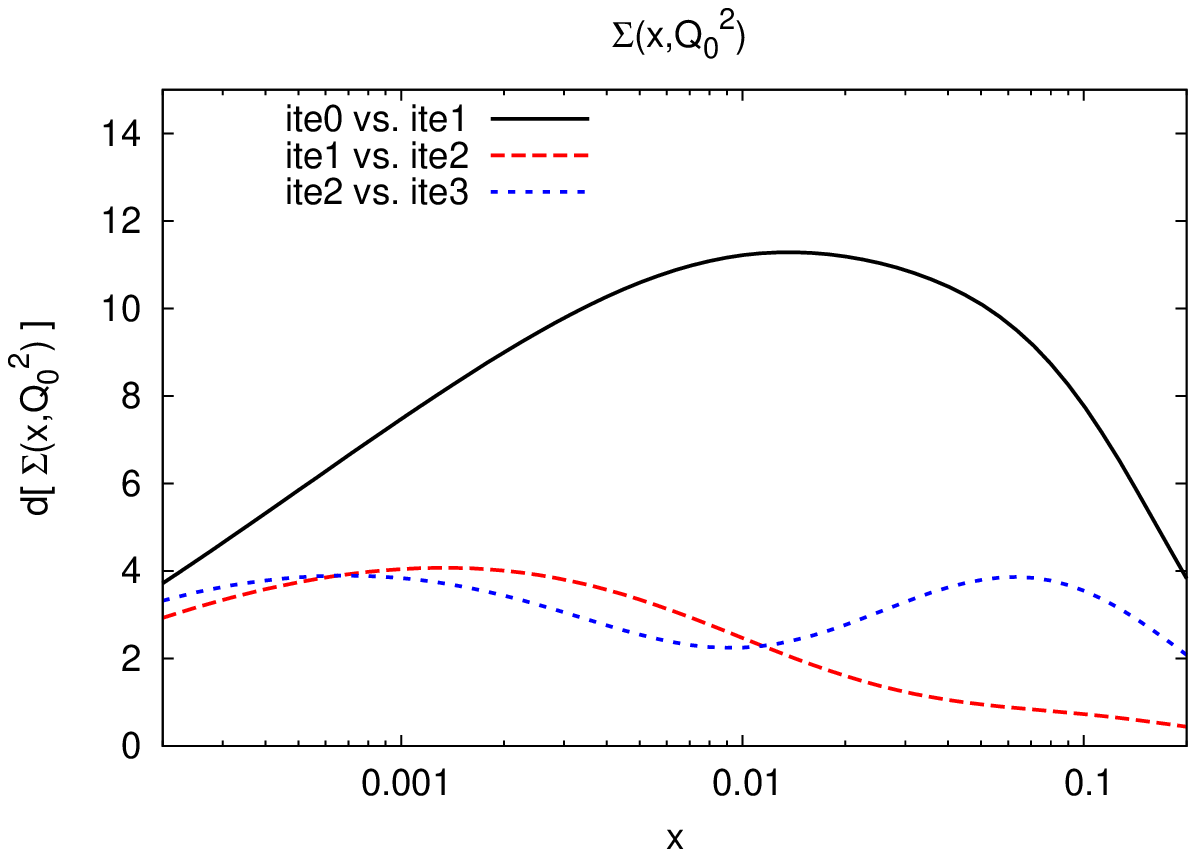}
\epsfig{width=0.49\textwidth,figure=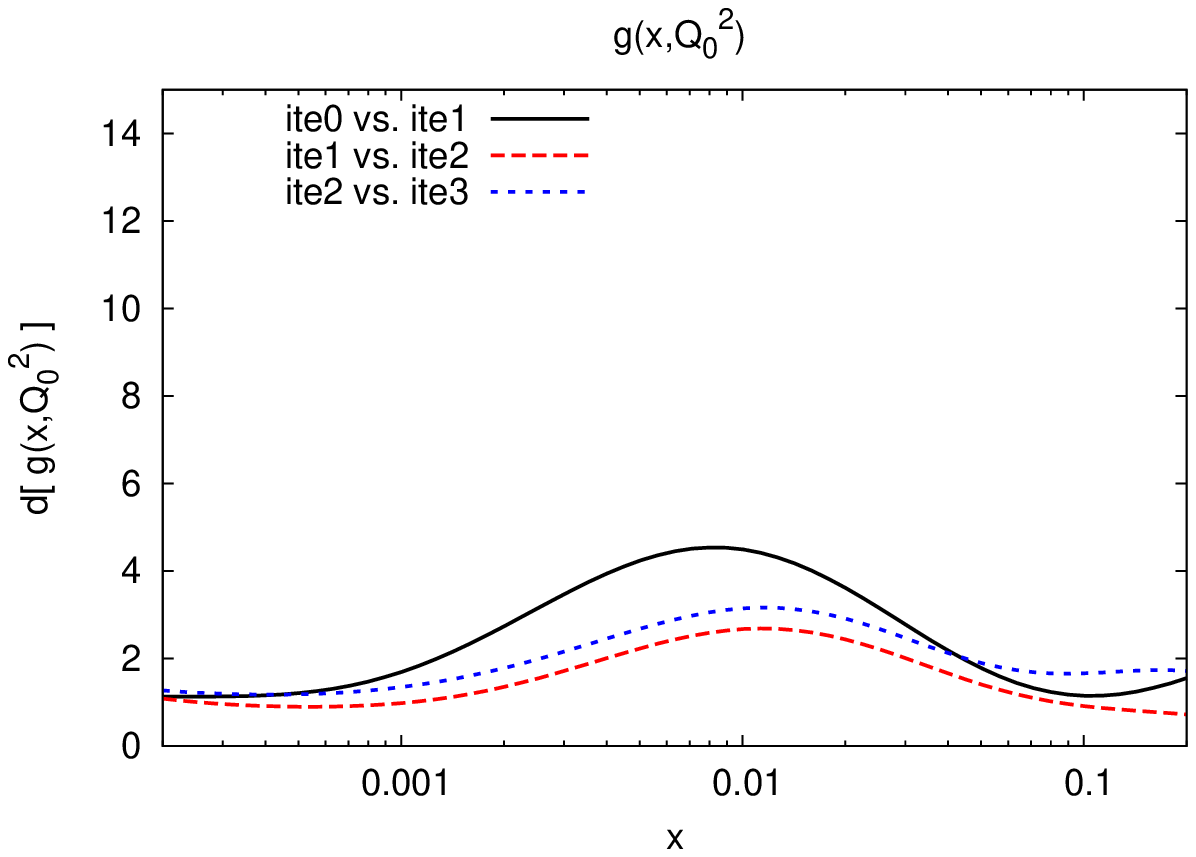}
\epsfig{width=0.49\textwidth,figure=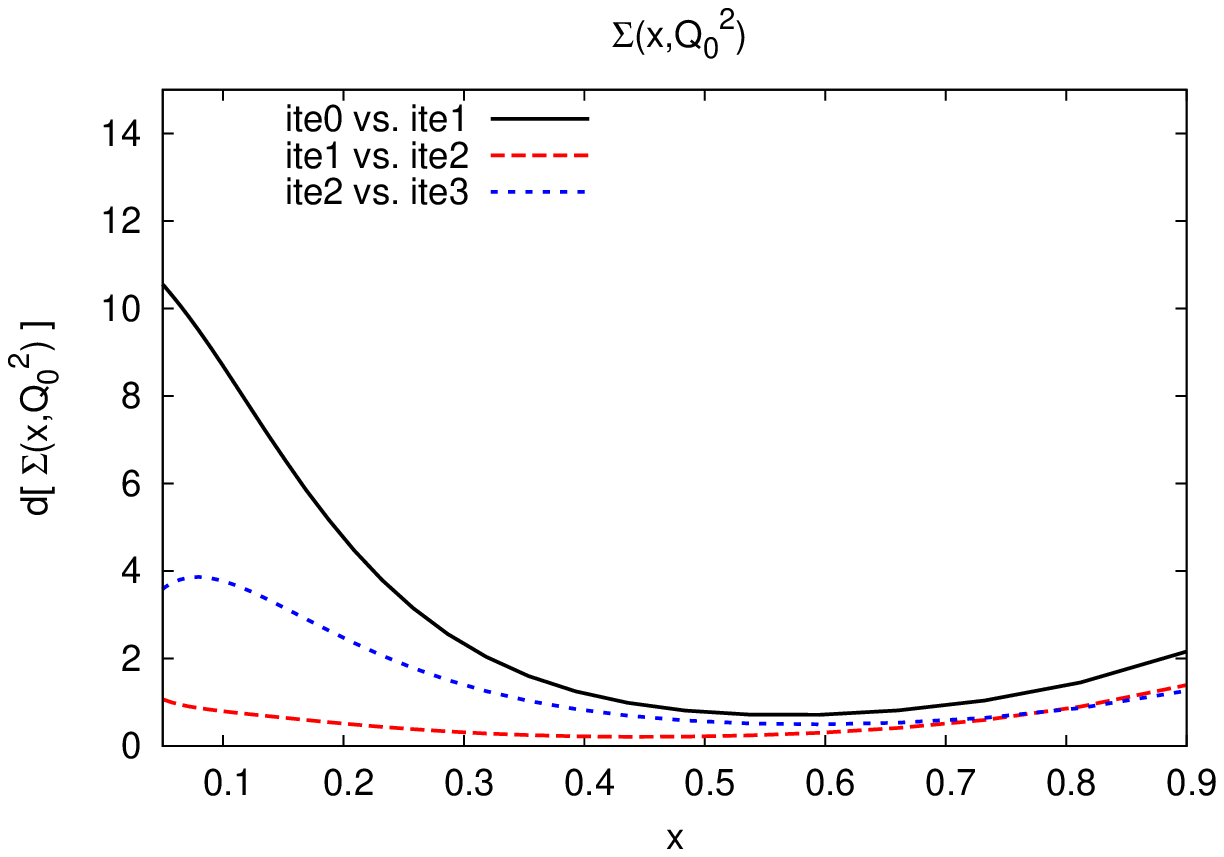}
\epsfig{width=0.49\textwidth,figure=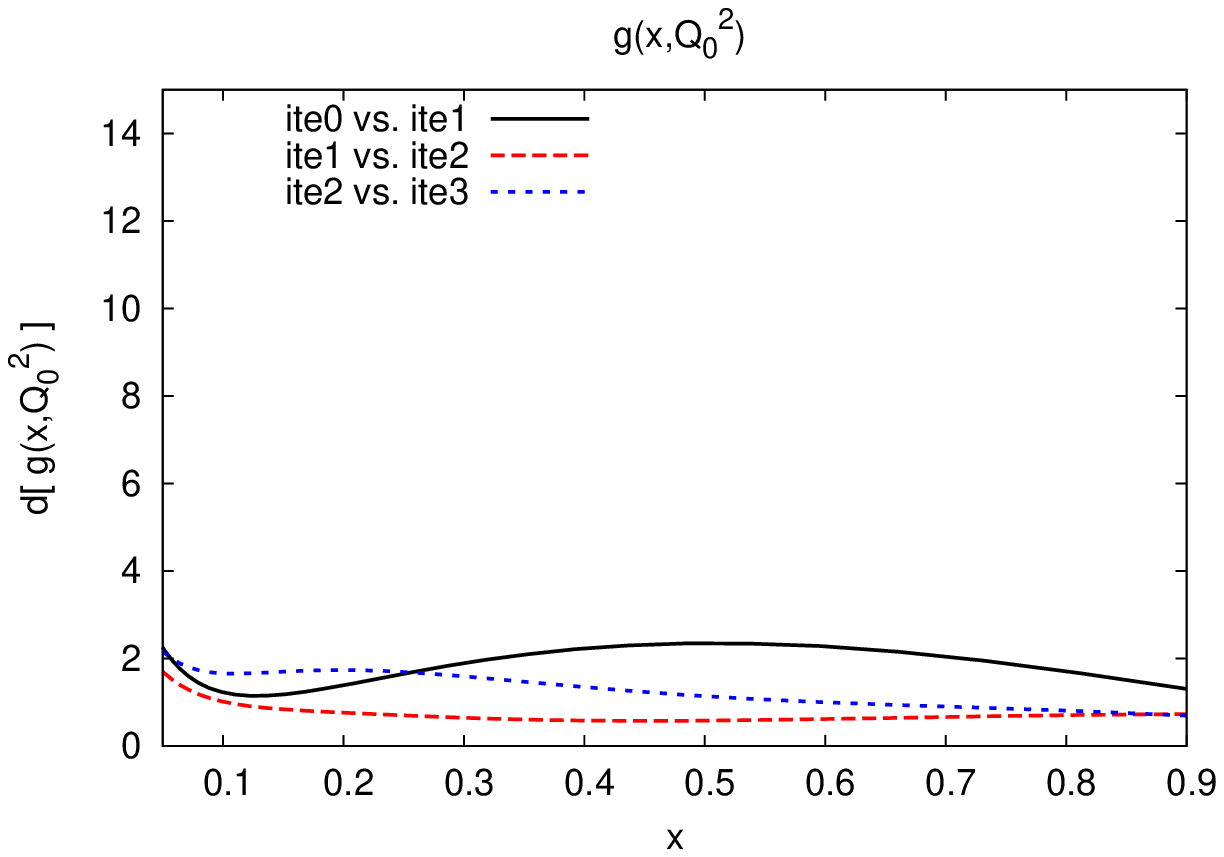}
\epsfig{width=0.49\textwidth,figure=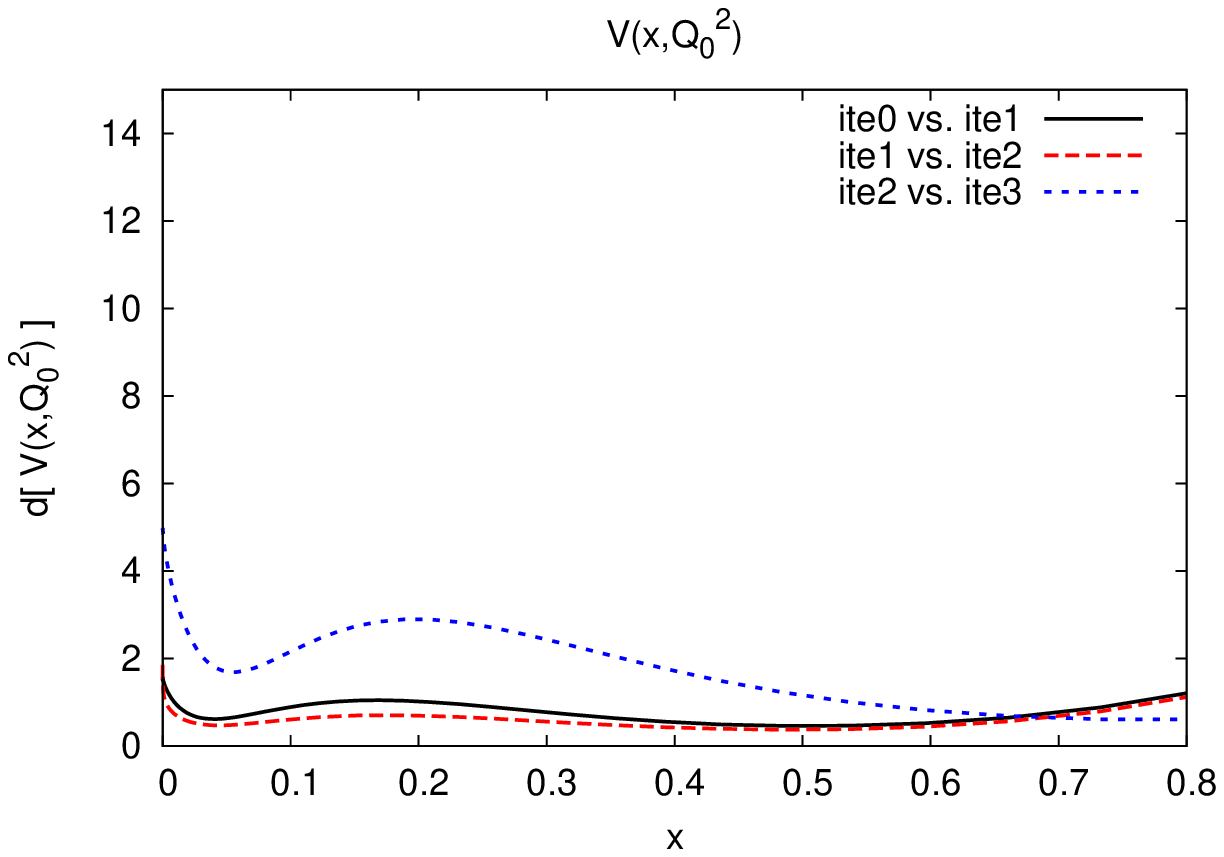}
\epsfig{width=0.49\textwidth,figure=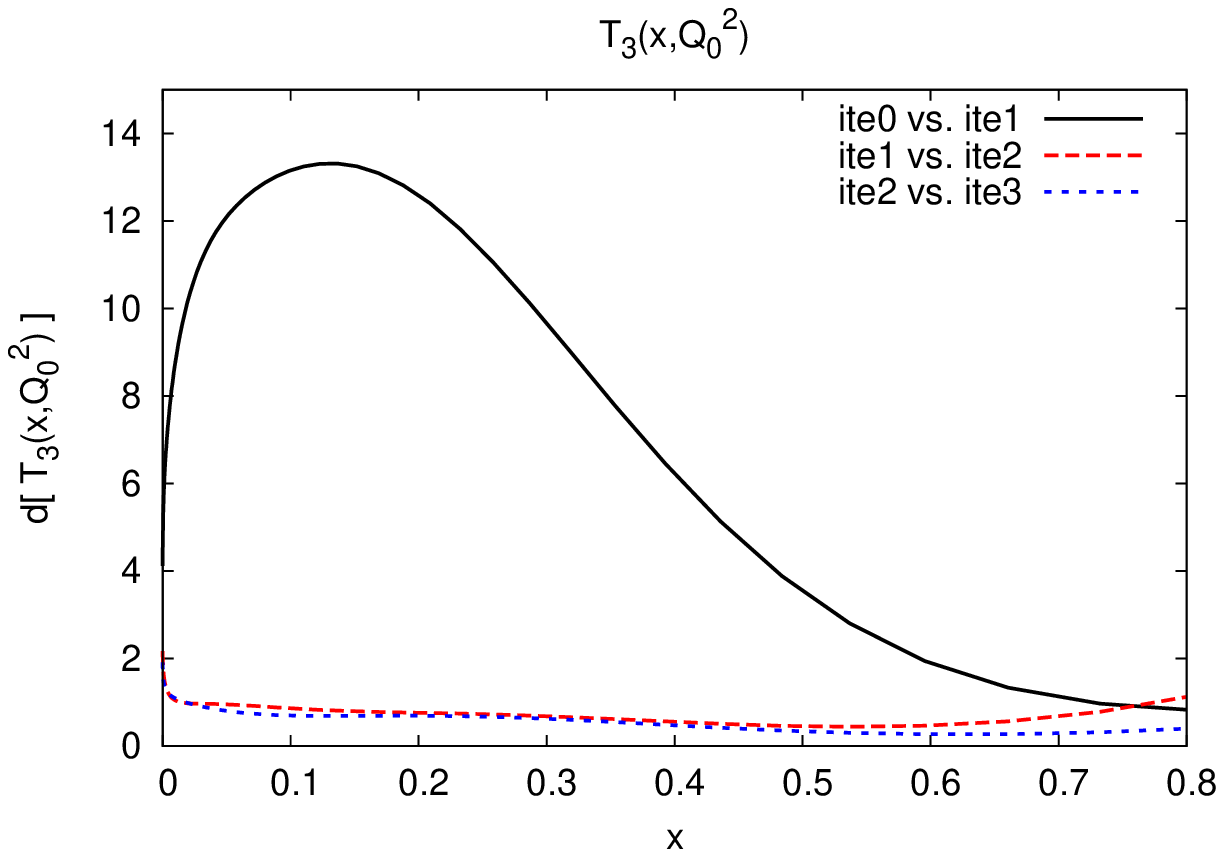}
\epsfig{width=0.49\textwidth,figure=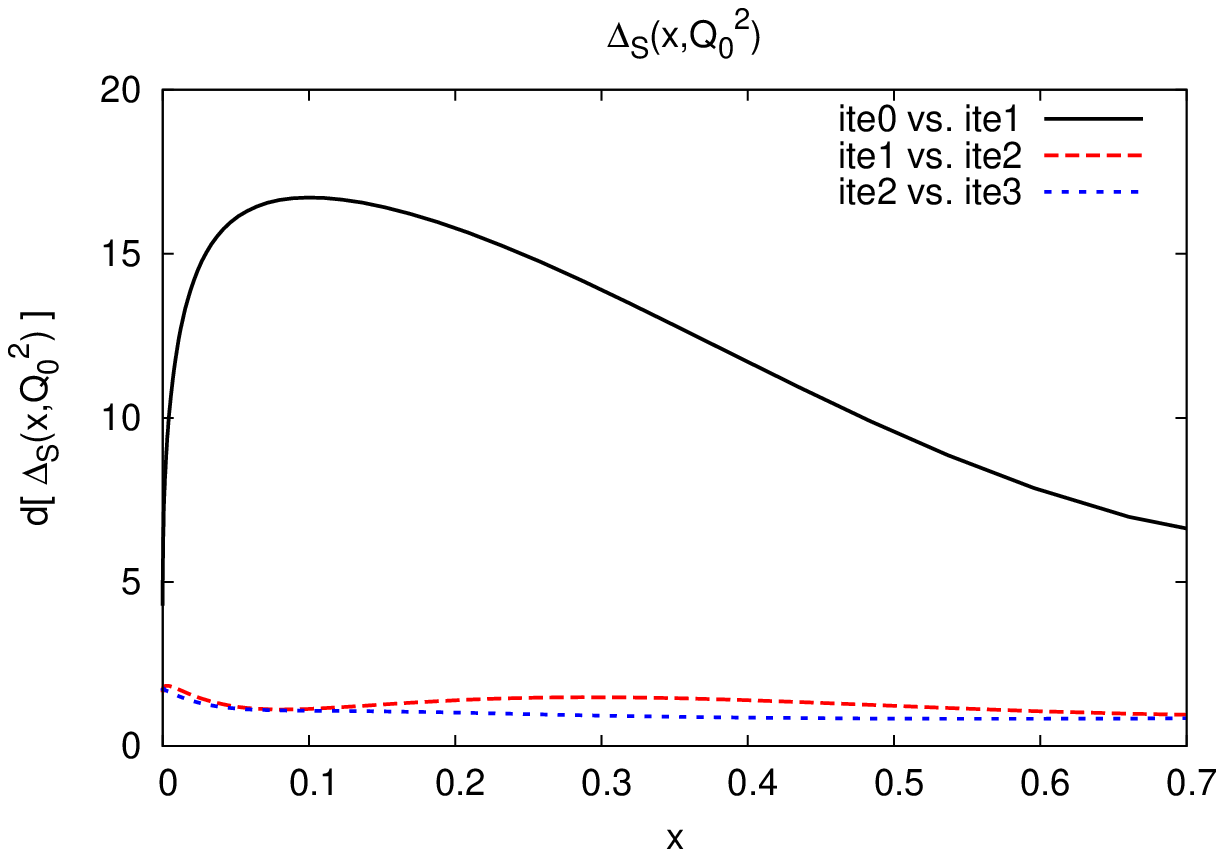}
\epsfig{width=0.49\textwidth,figure=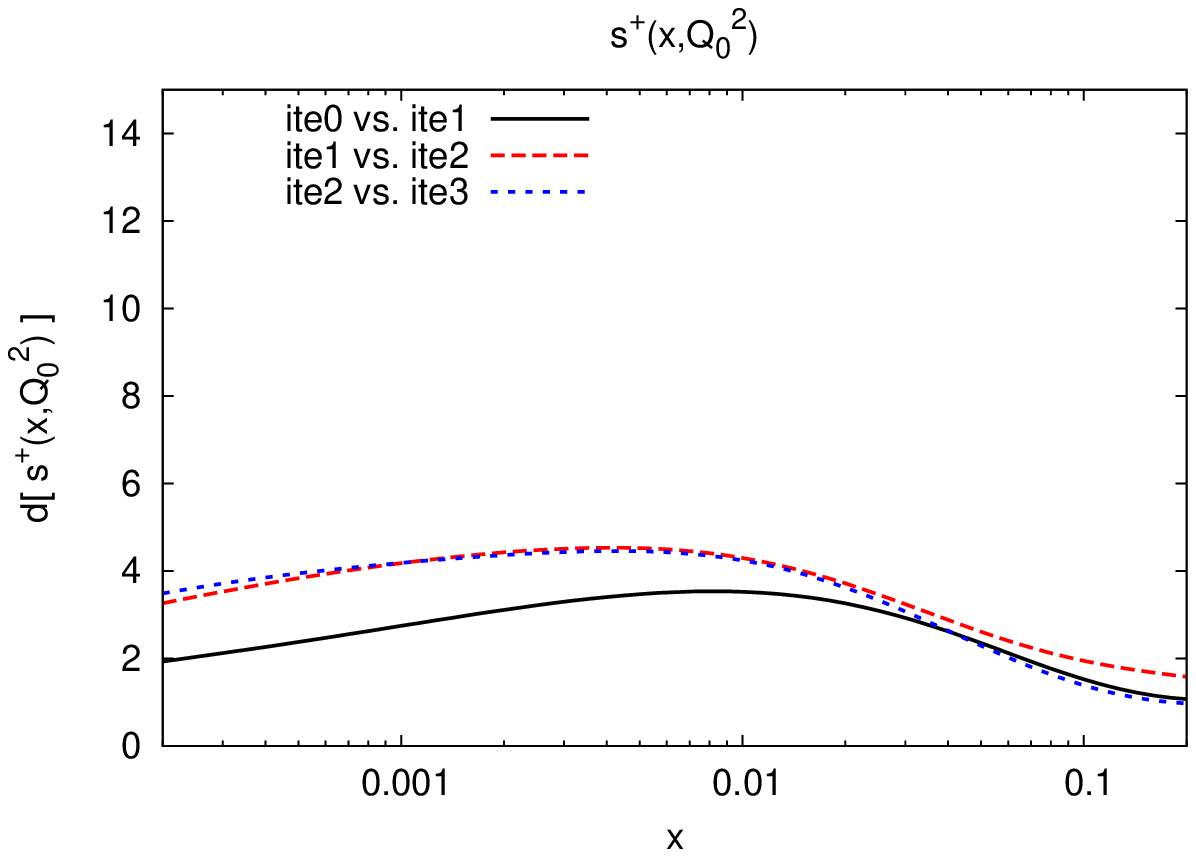}
\end{center}
\caption{\small \label{fig:distxplots-1000}
Distance between pairs of PDFs in subsequent iterations of the $t_0$ method.
For the singlet and the
gluon PDFs the distances are shown  
both at small $x$ and at large $x$.}
\end{figure}

These statements may be made more quantitative by looking at the distances 
between the various curves, in units of their standard deviations combined 
in quadrature (as defined in Appendix B of~\cite{NNPDFNS}). Distance
equal to one means that two curves are within one sigma of each other,
so the average distance between a random sample of curves should tend
roughly to one in the limit of large samples.
The distances
between ite1 and ite0, ite2 and ite1, and ite3 and ite2 are
shown as a function of $x$ in Fig.~\ref{fig:distxplots-1000} for
the various PDFs considered, and in Table~\ref{tab:stabtab2} averaged 
over $x$ (the average is performed by sampling the distance at ten
values of $x$ in the data region, see Ref.~\cite{NNPDF12}). 
Again the convergence at ite1 is apparent: only the distance 
between ite1 and ite0 is ever appreciably larger than unity. 
% ---------------------------------------------
\begin{table}
\begin{center}
{\footnotesize
\begin{tabular}{|c|c|c|c|c|}
\hline 
Experiment & {ite0} & {ite1} & {ite2} & 
{ite3}\\
\hline 
Total   & 1.32 & 1.25 & 1.25& 1.25 
\\
\hline
SLAC  & 1.33 & 1.30 & 1.31& 1.29
\\
\hline
BCDMS & 1.57 & 1.44 & 1.42 & 1.44
\\
\hline
NMC  & 1.71  & 1.66  & 1.67& 1.66
\\
\hline
NMC-pd  & 1.73  & 1.30 & 1.28& 1.29
\\
\hline
ZEUS & 1.05 & 1.02 & 1.03 & 1.03 
\\
\hline
H1  & 1.02 & 0.99  & 0.99& 1.00 
\\
\hline
CHORUS  & 1.38  & 1.31  & 1.32 & 1.33
\\
\hline
FLH108  & 1.65  & 1.67 & 1.67& 1.67 
\\
\hline
NuTeV Dimuon &  0.65 & 0.66 & 0.65& 0.68
\\
\hline
ZEUS HERA-II  & 1.53 & 1.49  & 1.49& 1.49
\\
\hline
\end{tabular}
}
\end{center}
\caption{\small The $\chi^2$ per degree of freedom, both total
and for individual experiments, for the various
iterations of the  NNPDF1.2 $t_0$ fits,
always with $1000$ replicas; ``ite0'' denotes the starting NNPDF1.2 fit.  
\label{tab:chi2tab2}
}
\end{table}
% ------------------------------------------------

It is clear from this analysis that the PDFs which are most affected by the
inclusion of normalization uncertainties in the fits are the 
singlet $\Sigma$ in the region $0.001 < x < 0.1$ (where there 
is tension between NMC and HERA), and the triplet $T_3$  and 
sea asymmetry $\Delta_S$ in the region $0.04< x <  0.4$ (due 
mainly to tension between BCDMS, NMC and CHORUS).
%------------------------------------------------------------
%------------------------------------------------------------

In Table~\ref{tab:chi2tab2} we show the value of
the $\chi^2$ per degree of freedom in the reference fit 
ite0 and in the various iterations
of the $t_0$ fit. We note the improvement in the fit quality from
the better handling of normalization uncertainties, particularly in BCDMS, 
NMC and CHORUS.  The table also shows that
the $\chi^2$ does not improve after the first iteration, which
provides further evidence
 for the convergence of the $t_0$ method at the first iteration.
%------------------------------------------------------------
\begin{figure}
\begin{center}
\epsfig{width=0.49\textwidth,figure=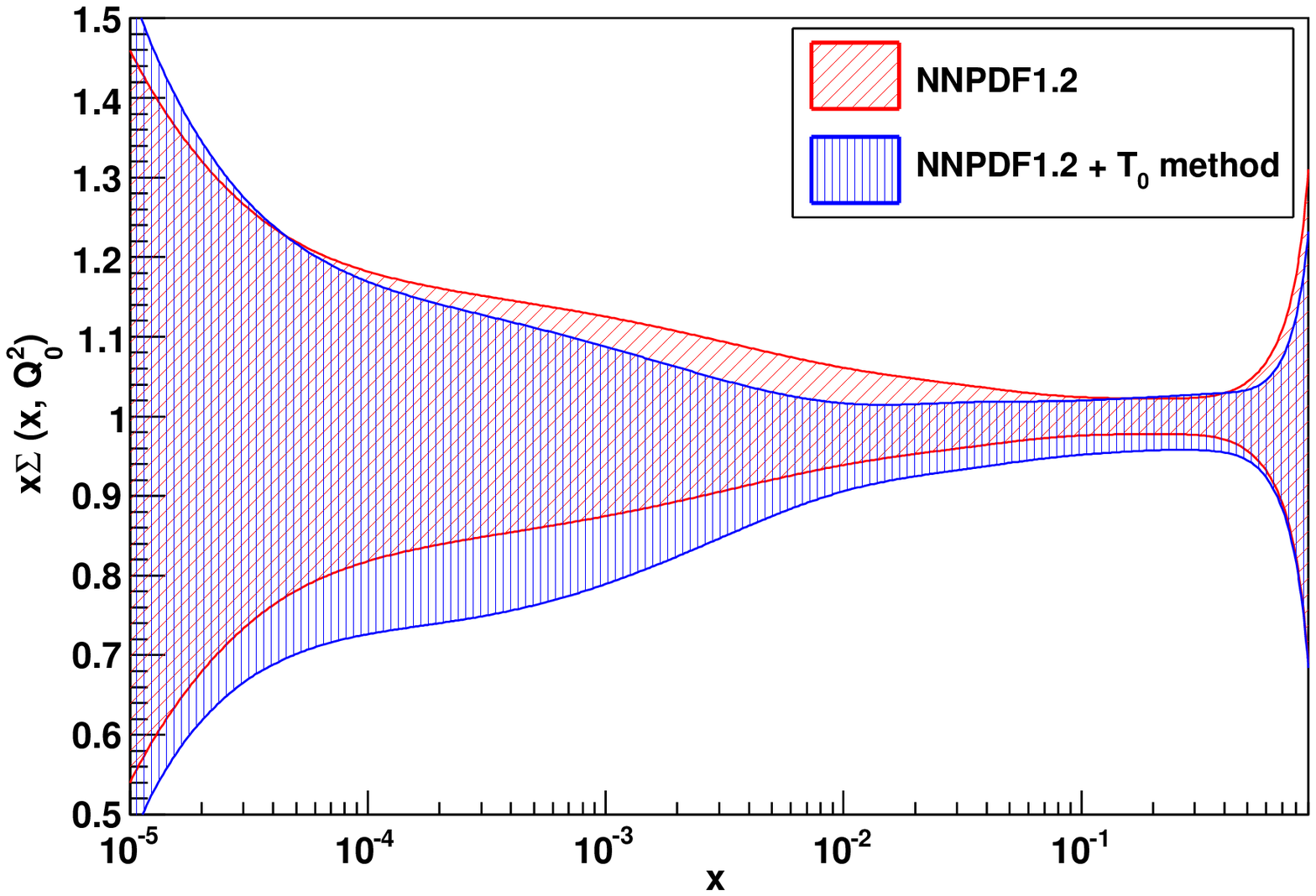}
\epsfig{width=0.49\textwidth,figure=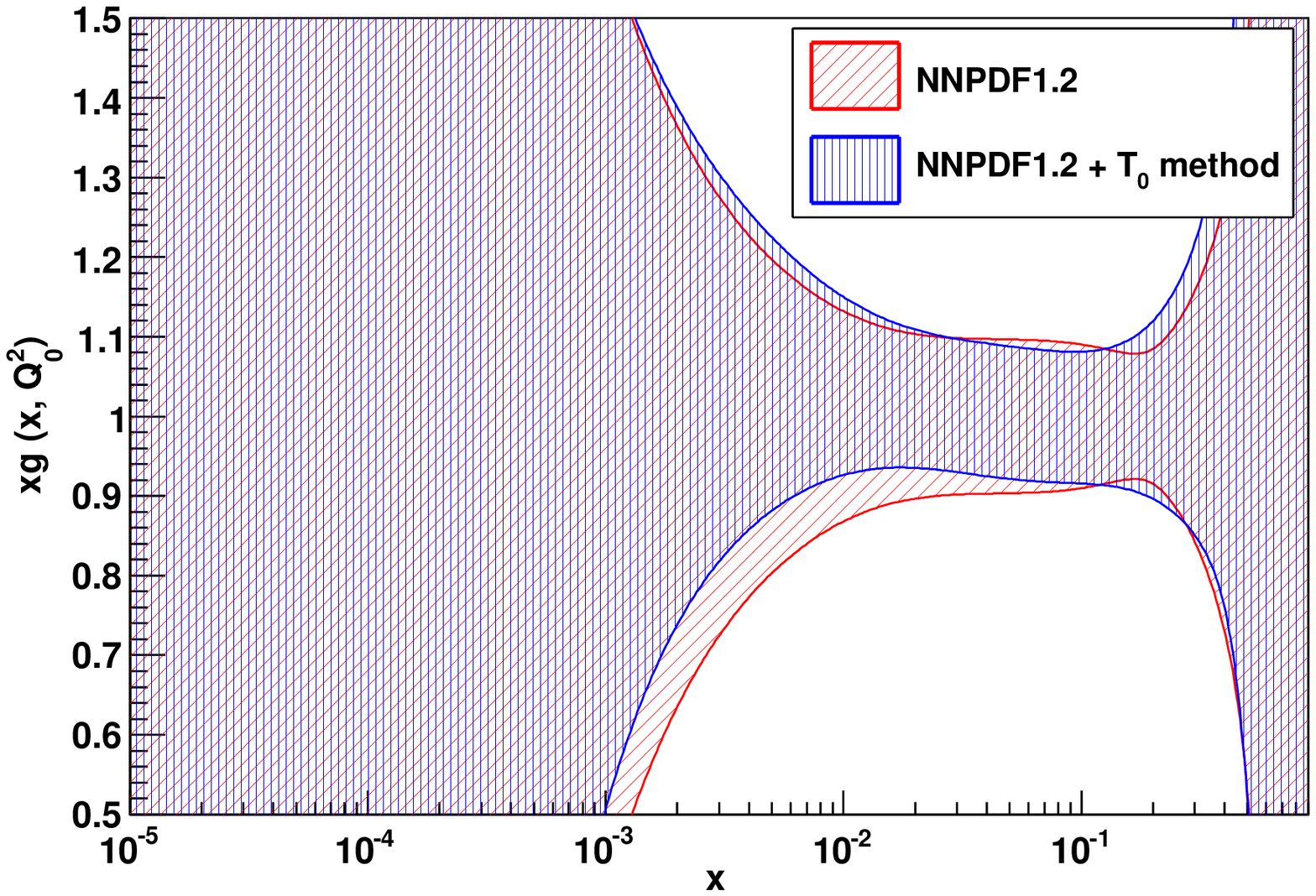}
\epsfig{width=0.49\textwidth,figure=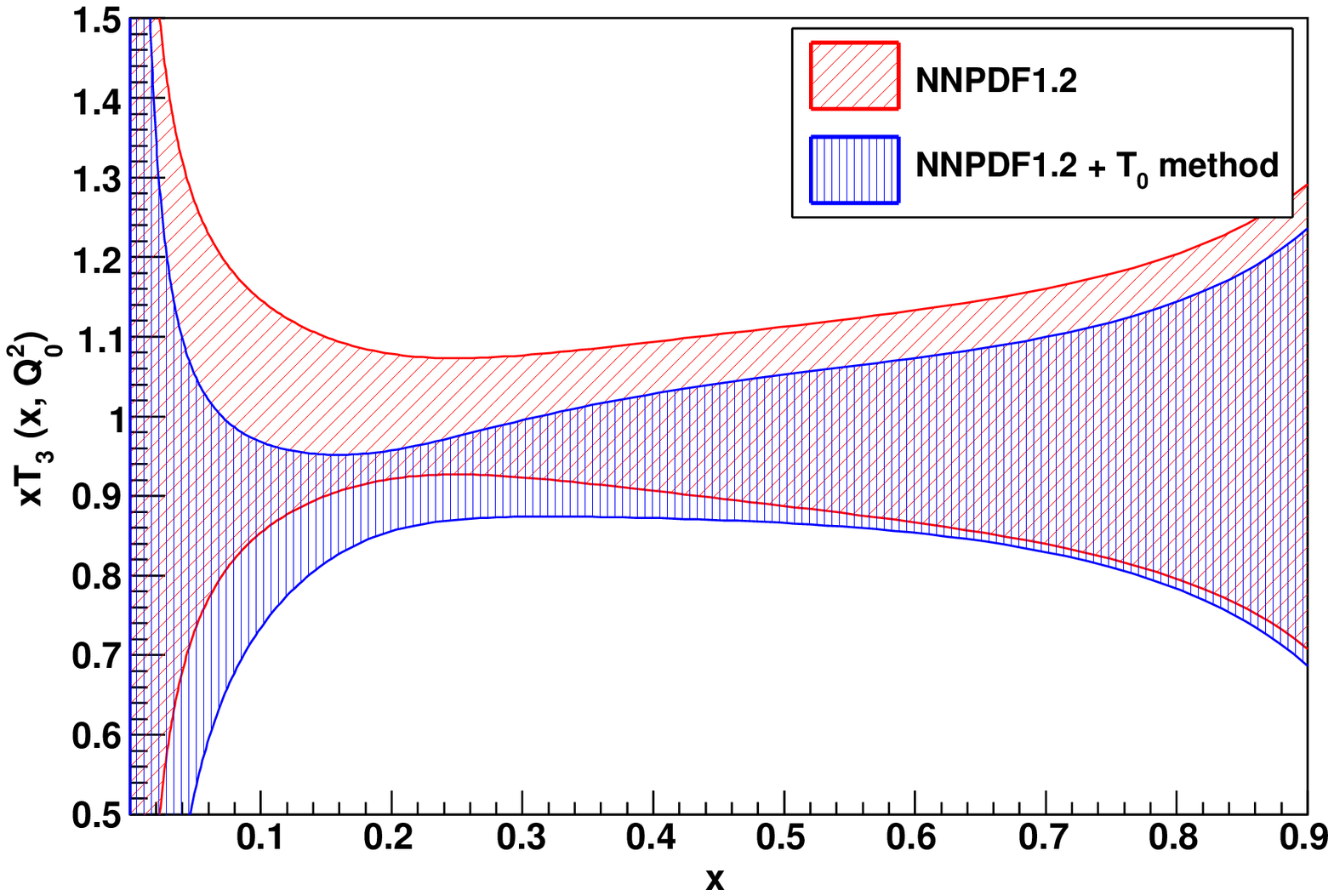}
\epsfig{width=0.49\textwidth,figure=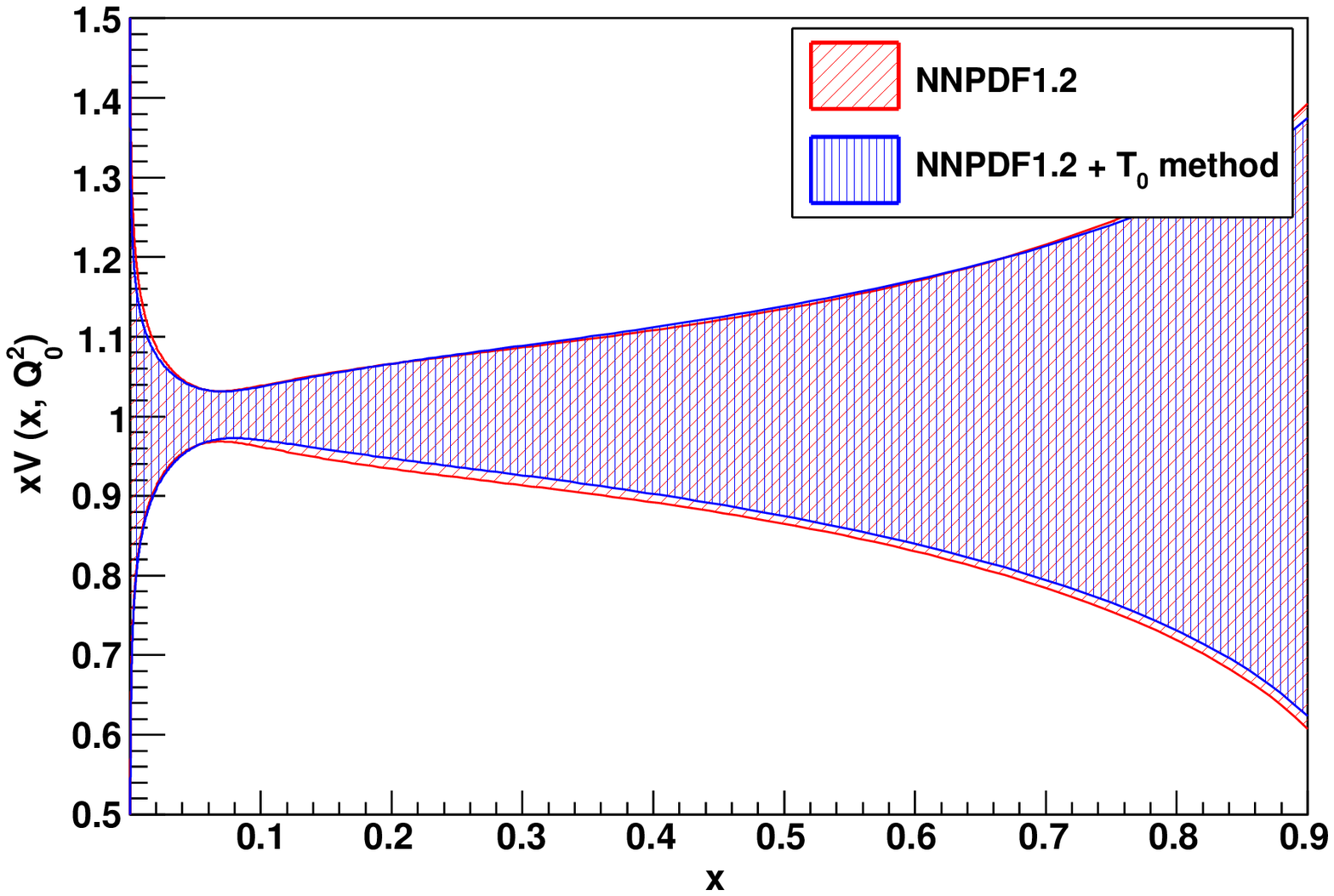}
\epsfig{width=0.49\textwidth,figure=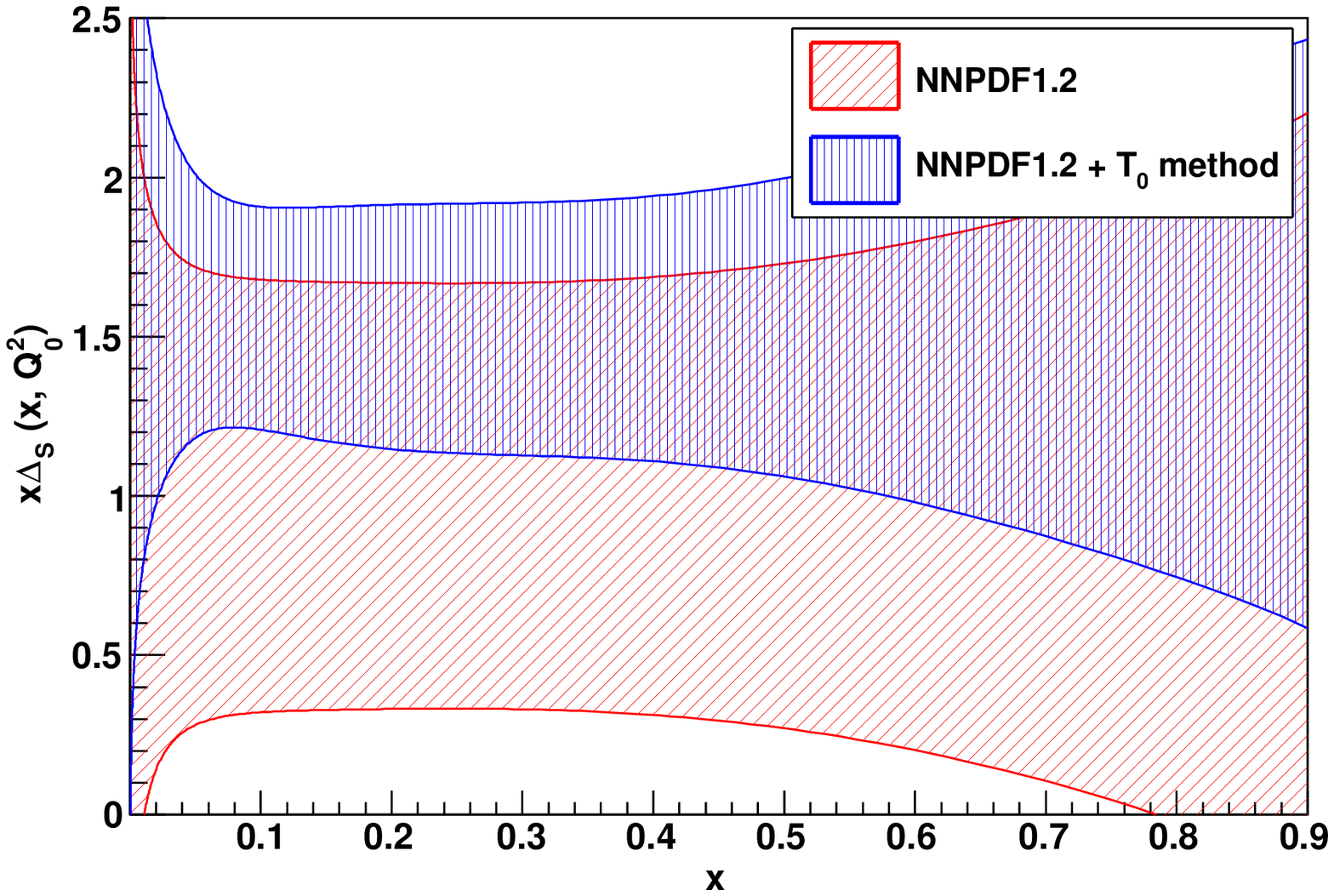}
\epsfig{width=0.49\textwidth,figure=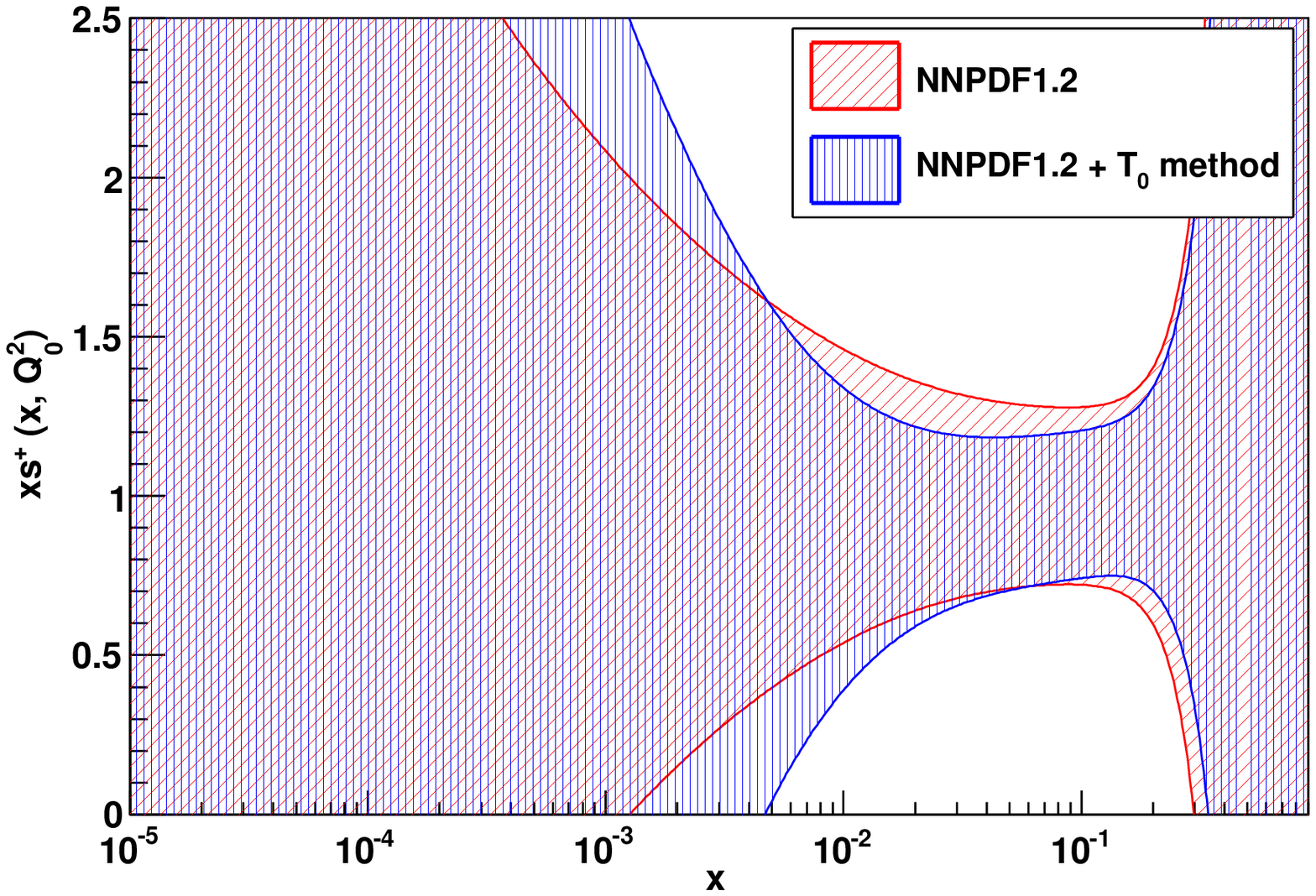}
\end{center}
\caption{\small \label{fig:pdfplots}
Comparison between the PDFs obtained from the {\bf ite3}
 NNPDF1.2$-t_0$ fit
with the standard ($t_0=0$)
NNPDF1.2~\cite{NNPDF12} fit. The quantity plotted is the ratio of the
difference between the  NNPDF1.2$-t_0$ and NNPDF1.2 results to the
NNPDF1.2 itself.}
\end{figure}
%------------------------------------------------------------

Finally, the PDFs from the starting $t_0=0$  NNPDF1.2  fit and those
of the  final
iteration (ite3) of the $t_0$ method, together with their respective
uncertainties, are compared
in Fig.~\ref{fig:pdfplots}. As expected from the distance
analysis the most important shifts in central values due to the 
inclusion of normalization uncertainties may be seen in the singlet, the 
triplet, and the sea asymmetry, though even these are generally small 
(ie around one-sigma). 
It is interesting to observe that these changes in central values 
of the triplet in the valence region, and of the sea asymmetry, are 
accompanied by a reduction in the overall uncertainty, due to the 
improved compatibility between the various datasets 
once normalization uncertainties are properly taken into account.

%\vfill\eject
\section{Conclusions}
\label{sec:con}

We have studied various methods for the inclusion of  multiplicative 
normalization uncertainties in combined fits to multiple data sets, using 
both Hessian and Monte Carlo methods, 
specifically but not necessarily in view of applications to PDF determination.
We reviewed how 
the simplest approach of including the normalization uncertainties in the
covariance matrix leads to the well--known ``d'Agostini bias'' \cite{dagos}, 
and showed 
that the commonly used penalty trick method, while fine for the analysis 
of a single set of experimental data, can lead to biases when combining 
several independent data sets. We then developed a new technique, the 
$t_0$-method, in which normalization uncertainties are introduced into 
the covariance matrix in such a way that the results are unbiased. While 
this technique requires iteration to self--consistency, we showed that 
in practice the convergence is very fast, so that only one iteration 
is generally required. 

To demonstrate the practical application of the $t_0$-method, we 
implemented it in the most recent published parton fit by the NNPDF 
collaboration, NNPDF1.2~\cite{NNPDF12}. This confirmed the rapid 
convergence of the technique, showed that the inclusion of normalization 
uncertainties can lead to a small improvement in the quality of the fit 
through the resolution of tensions between datasets, and moreover that 
where these tensions are significant this can lead to a  
subsequent reduction in PDF uncertainties.

We note that the $t_0$-method, while very well suited to the 
determination of PDF uncertainties by Monte Carlo methods, could 
also be used in the more traditional Hessian fitting methods, where it would 
lead to faster minimization (since in the $t_0$-method the dataset 
normalizations are not fitted), and more reliable central values (since 
unlike the penalty trick the $t_0$-method is free from systematic bias). 
However, the Hessian estimate of uncertainties is still  
a little less reliable than that from the Monte Carlo method, since 
quadratic cross-variance terms (such as in Eq.~(\ref{eq:varnm}))
are always missing, and Gaussian distributions of PDF parameters 
are always implicitly assumed.

The $t_0$-method has now been used in the global NNPDF
fits~\cite{NNPDF20}, where the compatibility of deep--inelastic and
hadronic data is a relevant issue. As expected, here the inclusion 
of the normalization uncertainties results in a significant improvement 
in the quality of the fit to the hadronic data sets.

Note that the d'Agostini and penalty trick biases discussed in this 
paper affect all multiplicative errors, not only overall normalizations. 
Many of the systematic errors in cross-section measurements are 
closer to multiplicative than additive (see for example Ref.~\cite{HERA}). 
The $t_0$-method might thus be developed into a general technique to 
obtain bias free fits to data sets with a variety of multiplicative 
systematic uncertainties.

%\vfill\eject

\section*{Acknowledgments}
We would like to thank R.S.~Thorne for raising the issue of
normalization uncertainties in the Hessian and Monte Carlo methods,
and for various discussions and correspondence. We also thank J.~Huston,
P.~Nadolsky and J.~Pumplin for discussions and correspondence.
On completing this work we discovered through
conversation with L.~Lyons that a similar method has been used to combine 
$B$ lifetime measurements~\cite{LMS}.  
This work was partly supported 
by the European network HEPTOOLS under contract MRTN-CT-2006-035505.

%\appendix
\end{document}